\documentclass{aa}
\usepackage{soul}   
\usepackage{graphicx}
\usepackage{amsmath}	
\usepackage{amssymb}	
\usepackage{rotating}
\usepackage{color}
\usepackage{lscape}

\usepackage{txfonts}
\begin{document}

    \title{Hunting Star-Forming Galaxies in the Gamma-Ray Domain}
   \subtitle{}

   \author{P. Kornecki\inst{1, 2}, J. Biteau\inst{3, 4}, C. Boisson\inst{1},  P. Cristofari\inst{1}}

   \institute{Laboratoire Univers et THeories (LUTH), Observatoire de Paris-Meudon, F-92195 Meudon Cedex \and
   Instituto de Astrofísica de Andalucía-CSIC, Glorieta de la Astronomía s/n, 18008, Granada, Spain
   \and Université Paris-Saclay, CNRS/IN2P3, IJCLab, 91405 Orsay, France \and Institut universitaire de France (IUF), France}

   \date{}
\authorrunning{P. Kornecki et. al}
 
  \abstract
   {Star-forming galaxies emit $\gamma$ rays with relatively low luminosity, but the study of their emission is no less captivating. While it is known that their $\gamma$-ray luminosity in the GeV band is strongly linked to their star formation, the origin of their emission at higher energies remains uncertain due to limited observations.}
   {Our aim is to assemble the largest possible sample of star-forming galaxies with potential detectability by the new-generation of Cherenkov telescopes.}
   {To achieve this, we compile a comprehensive sample of galaxies, including those previously detected by \textit{Fermi}-LAT in the GeV energy range as well as a larger sample of star-forming galaxies in the Local Volume that have been cataloged in the near-infrared band. We estimate their $\gamma$-ray flux assuming a proportional relationship with their star formation rate, and then select the brightest candidates. The predicted spectra in the TeV band are derived using a simple empirical model normalized to the star formation rate and a model based on extrapolating the latest \textit{Fermi}-LAT data to higher energies. The ground-based detectability of $\gamma$-ray emission from these sources is assessed through a comparison to the most recent instrument response functions.}
   {Our investigation reveals that almost a dozen star-forming galaxies may be detectable by  upcoming $\gamma$-ray telescopes.}
   {The observation of numerous star-forming galaxies in the TeV band is a fundamental piece of the panchromatic puzzle for understanding the physics inside these galaxies. The significant increase in the number of galaxies that can be studied in detail in the near future, particularly with the Cherenkov Telescope Array Observatory, promises a major step forward in the study of the conditions of acceleration and transport of cosmic rays in nearby extragalactic environments.}


\maketitle
%

\section{Introduction}

Star-forming galaxies (SFGs) radiate across a wide spectrum, from radio waves to $\gamma$ rays \citep{Condon1992,  Ackermann2012, Lacki2013}. Their non-thermal radio emission is caused by synchrotron radiation from energetic electrons in regions of intense star formation. These particles might originate from cosmic-ray (CR) accelerators, such as supernova remnants \citep{Axford1977, bell1978} or young massive stellar clusters \citep{Bykov1992}, which scale in number with the star formation rate (SFR) of their host galaxy. This assertion is supported by the robust correlation observed between the SFR and radio luminosity at 1.4\,GHz \citep{Condon1992, Yun2001, Bell2003, K22}.

Furthermore, CRs may interact with ambient protons in the interstellar medium, leading to the diffuse emission of high- and very-high-energy (VHE, $E > \mathrm{100}\,$GeV) $\gamma$ rays across the GeV and TeV bands \citep{Volk1996,Blom1999,Domingo2005,Persic2008,Yoast-Hull2013, Peretti2019, Krumholz2020, Roth2023}. To date, the \textit{Fermi} Large Area Telescope \citep[LAT;][]{Abdo2010} has detected over a dozen such galaxies in the $0.1-100$\,GeV energy range \citep{4fgl}. This diverse group of galaxies includes starburst galaxies (SBGs) and luminous to ultraluminous infrared galaxies (ULIRGs). Additionally, SBGs such as NGC\,1068 \citep{Antonucci1985} can host an inner Active Galactic Nucleus (AGN) on parsec scales. These non-jetted AGNs, often found in spiral galaxies, are classified as Seyfert nuclei.

Current research generally supports the idea that emissions from SFGs in the GeV range predominantly stem from hadronic processes \citep[e.g.,][]{Krumholz2020, Shimono2021, K22,Roth2023}. These processes may occur due to the interaction of energetic protons within the starburst nucleus on kpc scales or within star-formation environments of the surrounding galactic medium. The energy injected into CRs by highly active SFGs, such as Arp\,220, appears to be nearly entirely reprocessed into $\gamma$ rays. In contrast, a considerable portion of the CR energy might be dissipated by advection or diffusion in more quiescent SFGs \citep{K20, Werhahn2021, Roth2021, K22}. The injection and physical processes that CRs undergo in these galaxies are expected to determine the spectral slope in the GeV$-$TeV range. However, the transport mechanisms and the true properties of the CR spectrum in these galaxies remain under debate, partly because of the limited number of detections at the highest energies. 

Most SFGs observed at GeV energies with \textit{Fermi}-LAT have not been detected in the TeV regime, except for the two well-known SBGs, M82 and NGC\,253. These galaxies, being the closest and most active SBGs, have been detected in the VHE range by the VERITAS \citep{VERITAS2009} and H.E.S.S. observatories \citep{Abramowski2012, HESS2018}, respectively. Their emission in the TeV range is shown to be essentially an extension of the hadronic component observed in the GeV range by \textit{Fermi}-LAT. This provides evidence of proton acceleration up to multi-TeV energies in extragalactic star-forming environments.

A correlation between SFR and $\gamma$-ray luminosity, measured with \textit{Fermi}-LAT \citep{Ackermann2012, Ajello2020, K20}, has been observed in SFGs, supporting the connection between CRs and the star-formation process on galactic scales. However, this correlation in the GeV $\gamma$-ray range exhibits greater dispersion than the one found in the radio band at 1.4\,GHz \citep{Ackermann2012, K22}, which may arise from several factors. Some CR protons might lose their energy through non-radiative processes or escape the galaxy before radiating, leading to lower $\gamma$-ray emission levels than expected in the calorimetric limit of proton-proton processes \citep{Martin2014,Pfrommer2017,Peretti2019,Werhahn2021,Roth2021,Ambrosone2024}. Additionally, determining the location of the $\gamma$-ray emission within the galaxy is challenging due to the limited angular resolution of GeV$-$TeV instruments.

Alternative studies suggest that $\gamma$-ray emission from SBGs comes from the interaction of particles accelerated within galactic superwinds, driven by the combined effects of the stellar winds of young stars and supernova explosions. The accelerated particles then cool down typically a few  kpc away from the galactic disk \citep{Dorfi2012, Romero2018, Peretti2022}. Seyfert nuclei can also launch powerful winds \citep{Tombesi2010}, known as ultrafast outflows, that could accelerate CRs to hundreds of PeV \citep{Peretti2023}.
Investigations such as the studies of \citet{Lenain2010, Wojaczy2017, Lamastra2019} also argue that the non-thermal emission of some of these galaxies may be due to leptons accelerated as a result of AGN activity. Finally, a contribution to the diffuse $\gamma$-ray emission of SFGs is expected from non-resolved point sources such as Pulsar Wind Nebulae \citep[PWNs, see][]{Vecchiotti2022}, in particular in the TeV energy range \citep{Mannheim2012,Ohm2013,Do2021,Chen2024}. Consequently, the observed luminosity could be a combination of diffuse emission, AGN, population of PWNs, and emissions outside the galactic disk, e.g., from starburst superwinds.

The $\gamma$-ray energy range plays a pivotal role in elucidating the physical mechanisms governing the distribution of CRs within galaxies. Presently, alongside established observatories such as VERITAS, MAGIC, and H.E.S.S., a suite of new experiments are set to inaugurate a transformative era in $\gamma$-ray astronomy. Given the small number of SFGs detected at TeV energies by ground-based pointed telescopes, a priori promising instruments for establishing a larger population owing to their wide fields of view are The Large High Altitude Air Shower Observatory\footnote{http://english.ihep.cas.cn/lhaaso/?LMCL=i0dGTw} \citep[LHAASO,][]{LHAASO2019} in the northern hemisphere and the forthcoming Southern Wide-field Gamma-ray Observatory\footnote{https://www.swgo.org/SWGOWiki/doku.php} \citep[SWGO,][]{Albert2019}. The sensitivity of these observatories up to a few hundred TeV could open the door to the study of CR acceleration mechanisms and $\gamma$-ray absorption effects beyond 10\,TeV. The new-generation array of pointed $\gamma$-ray telescopes, the Cherenkov Telescope Array Observatory\footnote{https://www.ctao.org/} \citep[CTAO,][]{CTAO2023}, promises to revolutionize our understanding of SBGs with its extensive energy coverage from 20\,GeV to 300\,TeV, its exceptional sensitivity, and unparalleled angular resolution in the $\gamma$-ray band. The CTAO will help determine whether the observed $\gamma$ rays originate from diffuse emissions, similar to those detected at radio wavelengths, or from additional contributions such as AGNs or other point sources. Thus, the CTAO holds the potential to precisely delineate emission regions within some SBGs and elucidate the mechanisms governing CR transport.

The study of the most intense SFGs in the GeV range has been identified as one of the key science projects of the CTAO~\citep{CTABOOK2019}. Fainter $\gamma$-ray emitting galaxies, including those not yet detected by \textit{Fermi}-LAT, have so far been relatively neglected despite being crucial for establishing a relevant sample of SFGs at VHE. Several studies have attempted to evaluate the detectability of SFGs in the VHE range in recent years, albeit with a primary focus on the \textit{Fermi}-LAT energy range \citep{Shimono2021, Xiang2021}. These former studies did not aim for a census of SFGs observable at TeV energies.

In the present work, we investigate the chances of detection of SFGs in the TeV range, considering the up-to-date sensitivity of new-generation ground-based telescopes. Our analysis incorporates recent \textit{Fermi}-LAT data and integrates information from full-sky optical and infrared surveys mapping galaxies in the nearby universe. To evaluate their observability, we also consider their sky positions to select the corresponding arrays for visibility and the zenith angle to choose the most accurate response available.
We work under the assumption that the GeV emission is coming mainly from proton-proton interactions and that the VHE spectrum is an extension of the emission observed in the GeV range. Our study encompasses two groups of SFGs: those previously detected with \textit{Fermi}-LAT in the GeV range, and a selection of galaxies that have not been detected at high energies so far. In Sect.~\ref{sec:GeVcandidates}, we update the study of the correlation between SFR and $\gamma$-ray luminosity for GeV-detected galaxies and search for such emission in a larger sample of SFGs. In Sect.~\ref{sec:TeVcandidates}, we evaluate the detectability at VHE of all the galaxies in the sample and provide a list of the best candidates to be observed with current and upcoming observatories. Our conclusions are outlined in Sect.~\ref{sec: conc}.

\section{Candidate SFGs in the GeV energy range}
\label{sec:GeVcandidates}

\subsection{Sample of galaxies of interest}
\label{sec: nonGeVDet}

We explore the prospect of discovering new SFGs in the GeV energy range and subsequently assess their potential detection at TeV energies. To this end, we employ the Revised MANGROVE Sample from \cite{Biteau2021}, which represents a comprehensive collection of SFGs over more than 90\% of the sky and out to 350\,Mpc ($z<0.08$). This near-infrared flux-limited sample comprises around 400,000 galaxies, the distances of half of which are estimated photometrically and half spectroscopically (or from the cosmic distance ladder for the closest galaxies). Stellar masses are estimated from emission in the W1 band of the Wide-field Infrared Survey Explorer \citep[WISE,][]{WISE2010}, using the mass-to-light ratio inferred for a Chabrier initial mass function \citep{Chabrier2003}. Galaxies with strong near-infrared emission associated with an AGN were excluded from the initial MANGROVE based on their W1-W2 color \citep{2020MNRAS.492.4768D}. The SFR of the galaxies kept in the sample is estimated from their H$\alpha$ emission for galaxies in the Local Volume ($d < 11\,$Mpc) studied by \cite{2018MNRAS.479.4136K}. These authors accounted for dust attenuation through the apparent orientation of the galaxy. For other galaxies, scaling relations  are used by \cite{Biteau2021} to infer SFR from stellar mass, albeit with a dispersion ranging from 0.3\,dex to 0.8\,dex depending on the morphological class of the galaxy.

\begin{table*}
\centering
\caption{\label{tab: nonGeVdet} Sample of SFGs not included in the 4FGL-DR4 that satisfy the selection criteria defined in Sec.~\ref{subsec: nonGeVDet_sel_crit}.}
\begin{tabular}{lccccccc}
\hline\hline

Name 	&	 $d$ 	&	 R.A. 	&	 Dec 	&	 $\dot{M}_*$ 	&	 TS	&	 $\mathcal{F}^\mathrm{UL}_\mathrm{0.1-100\,GeV}$ 	&	 $ \mathcal{F}_\mathrm{2\,GeV}^{\mathrm{UL}}$  \\
 	&	 $\mathrm{Mpc}$ 	&	  deg 	&	 deg 	&	 $M_{\odot}\,\mathrm{yr^{-1}}$ 	&	 	&	 $\mathrm{ 10^{-10} \, GeV \, cm^{-2} \, s^{-1}}$ 	&	 $ 10^{-14}\, \mathrm{erg \, cm^{-2} \, s^{-1}} $  	\\
\hline															
NGC\,6822 	&	 $0.4603 \pm 0.0040$ 	&	296.24	&	-14.80	&	 $0.0070 \pm 0.0006$\tablefootmark{d}  	&	 8	&	8.72	&	20.5  \\             
M33     	&	 $0.853 \pm 0.016$   	&	 \phantom{0}23.46 	&	 \phantom{-}30.66 	&	 $0.251 \pm 0.013$\tablefootmark{b}  	&	22	&	10.6	&	24.9 \\             
NGC\,300 	&	 $1.944 \pm 0.040$  	&	 \phantom{0}13.72 	&	-37.68	&	 $0.138 \pm 0.005$\tablefootmark{a}  	&	 0	&	0.63	&	1.48 \\             
NGC\,55 	&	 $1.981 \pm 0.020$   	&	 \phantom{00}3.79 	&	-39.22	&	 $0.186 \pm 0.010$\tablefootmark{b}  	&	 0	&	1.56	&	3.66 \\             
IC\,342 	&	 $2.24 \pm 0.10$   	&	 \phantom{0}56.70 	&	 \phantom{-}68.10 	&	 $0.40 \pm 0.03$\tablefootmark{c}  	&	 3	&	3.77	&	8.86 \\             
NGC\,1569 	&	 $3.25 \pm 0.24$ 	&	 \phantom{0}67.70 	&	 \phantom{-}64.85 	&	 $0.70 \pm 0.08$\tablefootmark{a}  	&	 10	&	9.93	&	23.4 \\             
M81     	&	 $3.61 \pm 0.20$   	&	148.89	&	 \phantom{-}69.07 	&	 $0.42 \pm 0.04$\tablefootmark{b}  	&	 0	&	1.37	&	3.21 \\             
M83     	&	 $4.79 \pm 0.09$   	&	204.25	&	-29.87	&	 $3.29 \pm 0.10$\tablefootmark{a}  	&	18	&	9.66	&	22.7 \\             
NGC\,3621 	&	 $6.70 \pm 0.37$ 	&	169.57	&	-32.81	&	 $0.86 \pm 0.07$\tablefootmark{a}  	&	 7	&	7.31	&	17.2 \\             
M101    	&	 $6.76 \pm 0.11$   	&	210.80	&	 \phantom{-}54.35 	&	 $3.13 \pm 0.22$\tablefootmark{b}  	&	 3	&	3.13	&	7.37 \\             
NGC\,6946 	&	 $6.95 \pm 0.38$ 	&	308.71	&	 \phantom{-}60.15 	&	 $3.70 \pm 0.33$\tablefootmark{c}  	&	 0	&	1.52	&	3.56 \\             
M106    	&	 $7.54 \pm 0.09$   	&	184.74	&	 \phantom{-}47.30 	&	 $1.03 \pm 0.08$\tablefootmark{d}  	&	 13	&	3.32	&	7.80 \\             
M51     	&	 $8.34 \pm 0.30$   	&	202.47	&	 \phantom{-}47.23 	&	 $2.83 \pm 0.15$\tablefootmark{a}  	&	 7	&	4.82	&	11.3 \\             
\hline\hline
\end{tabular}

\tablefoot{
The columns $d$, R.A. and Dec are the distance, right ascension and declination in ICRS reference system of the galaxy center taken from SIMBAD astronomical database. $\dot{M}_*$ is the SFR. TS is the obtained test statistic for detection in the GeV energy range. $\mathcal{F}^\mathrm{UL}_\mathrm{0.1-100\,GeV}$ and $ \mathcal{F}_\mathrm{2\,GeV}^{\mathrm{UL}}$ are the 95$\%$ confidence-level upper limits on the energy flux integrated over $0.1 - 100$\,GeV and at 2\,GeV, respectively, assuming a fixed spectral index of 2.2.}

\tablebib{
\tablefoottext{a}{Computed from FUV \citep{GildePaz2007} + IRAS $25\, \mu\mathrm{m}$ \citep{Sanders2003} fluxes.}
\tablefoottext{b}{Computed from FUV \citep{GildePaz2007} + IRAS $25\, \mu\mathrm{m}$ \citep{rice1988} fluxes.}
\tablefoottext{c}{Computed from H$\alpha$ \citep{Kennicutt2008} + IRAS $25\, \mu\mathrm{m}$  \citep{Sanders2003} fluxes.}
\tablefoottext{d}{Computed from H$\alpha$ \citep{Kennicutt2008} + IRAS $25\, \mu\mathrm{m}$  \citep{rice1988} fluxes.}}
\end{table*}

\label{subsec: nonGeVDet_sel_crit}

Our sample is composed of galaxies with a $\gamma$-ray emission from star formation that may be challenging to detect at TeV energies, either due to low SFR or large distance from Earth. 
To ensure the relevance of the subsample to be examined, we first selected galaxies that are a priori bright enough to be detectable by new-generation TeV $\gamma$-ray observatories, as determined by the ratio $\text{SFR}/4\pi d^2$. The new-generation observatories target sources with fluxes down to ${\sim}\, 1\,$mCrab in ${\sim}\, 50\,$h, to compare with the flux of the SBG NGC\,253 of about ${\sim}\, 3.5\,$ mCrab at 1\,TeV, as measured with previous generation observatories in 160\,hours of observations \citep{HESS2018}. We thus select all galaxies in the Revised MANGROVE sample that satisfy the relation $\text{SFR}/4\pi d^2 > \alpha \text{SFR}_\text{NGC\,253}/4\pi d_\text{NGC\,253}^2$, where $\alpha = 1/3.5$. To ensure the completeness of the sample, we applied the same selection criterion to galaxies observed in the infrared with IRAS \citep{Sanders2003} and selected by the \textit{Fermi}-LAT Collaboration \citep{Ackermann2012}, using their radio, infrared, and HCN ($J=1-0$) luminosities as a rough tracer of the SFR. This verification, illustrated in Appendix~\ref{app: summary}, confirms the greater completeness of the Revised MANGROVE sample in the Local Volume and allows us to include NGC\,1068, otherwise excluded due to its powerful Seyfert nucleus. We exclude galaxies with dwarf elliptical, lenticular, or elliptical  morphology, namely the Sagittarius Dwarf Galaxy, Maffei\,1, Centaurus\,A, and M49. After this exclusion, our sample comprises 21 galaxies, including 8 already detected by \textit{Fermi}-LAT (LMC, SMC, NGC\,253, M82, NGC\,4945, NGC\,2403, Circinus, M31). Table~\ref{tab: nonGeVdet} lists the remaining 13 galaxies, which have not been yet detected by \textit{Fermi}-LAT.

We collected the luminosity distance and associated uncertainty for each galaxy from the Extragalactic Distance Database,\footnote{\url{https://edd.ifa.hawaii.edu/dfirst.php?}} based on the recent update of Cosmicflows-4 \citep{Tully2023}. For SFR values, discrepancies arise when they are derived from different spectral indicators (such as UV, H$\alpha$, [OII] lines, and far-infrared fluxes). Many of these disparities are primarily attributed to uncertainties on dust attenuation. To address this issue, we refined the estimation of the SFR following \cite{K20} and employed multi-wavelength tracers \citep[in particular FUV+IR$_{25 \mu \rm{m}}$ or H$\alpha$+IR$_{25 \mu \rm{m}}$ and total IR emission in 8--1000\,$\mu$m,][]{kennikutt1998A, Kennicutt2012}. We took the IR$_{25 \mu \rm{m}}$ fluxes from \cite{Sanders2003} and \citet{rice1988}, FUV from \cite{Cortese2012} and \cite{GildePaz2007}, and H$\alpha$ from \cite{Kennicutt2008}. For the nine galaxies with data available in both FUV and H$\alpha$, we checked that the two SFR estimates align within measurement uncertainties. In such cases, we chose to use the FUV values for further analysis. We verified that the refined estimates of the SFR are statistically consistent with the initial estimate from the revised MANGROVE sample and have lower uncertainties. The uncertainties in the quoted SFR values account only for the uncertainties in the distances, FUV/H$\alpha$, and IR flux measurements, and not for uncertainties in the calibration constants nor for dispersion around the calibration relations. The estimated distances and SFRs are listed in Table~\ref{tab: nonGeVdet} together with the relevant references.

\subsection{Fermi-LAT Data Analysis for non GeV-detected SFGs}

\label{sec: GeVana}

To search for high-energy $\gamma$-ray emission associated with galaxies not included in the 4FGL, we use \textit{Fermi}-LAT PASS~8 archival data,\footnote{https://fermi.gsfc.nasa.gov/cgi-bin/ssc/LAT/LATDataQuery.cgi} spanning almost 15 years of observations from 2008 August 4 to 2023 July 12. 
For each galaxy, we downloaded $\gamma$-ray events in a region of interest (ROI) of 15-deg radius, centered on the coordinates from Simbad.\footnote{http://simbad.u-strasbg.fr/simbad/} The data in the energy range $0.1-100$\,GeV were analyzed using the publicly available Fermi Science Tools v2.2 interfaced with Fermipy v1.2.0 \citep{Wood+2017} in combination with the latest IRFs (P8R3$\_$SOURCE$\_$V3).

We adopted a source-type event selection (evclass=128, evtype=3) and applied standard data-quality selection criteria (``DATA$\_$QUAL$>$~0~$\&\&$~LAT$\_$CONFIG==1''). Zenith angles are limited to 90\,deg to minimize contamination from the Earth limb. We binned spatially the data  with a scale of 0.1\,deg per pixel and used 8 logarithmically-spaced bins per energy decade. A model file including all sources of the 4FGL-DR4 catalog that fall in the ROI as well as the Galactic diffuse emission and extragalactic isotropic emission components was build. A standard binned likelihood analysis was applied in an iterative way. We fixed the spectral-shape parameters of 4FGL-DR4 sources more than 10\,deg away from the galaxy to account for event leakage in the ROI due to the larger point spread function at lower energies. In a second step, the sources contributing to less than 5\% of the total number of counts in the ROI ($N_\mathrm{pred}/N_\mathrm{count}  < 0.05$)  and/or with low significance, $\mathrm{TS} < 9$, had their parameters frozen. The detection significance of each source is evaluated with a Test Statistic (TS) defined as TS = 2 $\Delta$log(likelihood) between models with and without the source. In the end, the only free parameters are those of sources less than 3\,deg away from the target, if not frozen in the previous step, and the normalizations of the two diffuse background components. After the fit, the residuals and TS maps were obtained by subtracting the best-fit model from the data. Whether or not an excess is found in the TS map at the source location, in order to check for any detection, we refit the data adding a new point source at the galaxy position, leaving free the photon index and normalization of the source of interest. 

All the galaxies show significance considerably lower than $5\sigma$ ($\mathrm{TS} < 25$) when the photon index is left free, except for M83 and M33 with TS values of 18 and 22, respectively.  Notably, the excesses from M83 and M33 correspond to photon indices of $2.0 \pm 0.2$, and $2.3 \pm 0.2$, respectively, in line with the expected value for a SFG.

We calculated upper limits on the flux for each galaxy assuming a fixed photon index of $2.2$ for each source. This value of the photon index is expected in theoretical models describing the diffuse $\gamma$-ray emission from these galaxies, under the assumption that the emission mainly comes from proton-proton interactions \citep{Yoast-Hull2013, Wang2018,Peretti2019,K20}. We provide 95 $\%$ confidence-level upper limits on the energy flux integrated over the entire energy range of the analysis ($\mathcal{F}^\mathrm{UL}_\mathrm{0.1-100\,GeV}$ in GeV$\,\rm{cm}^{-2}\,s^{-1}$), as shown in Table~\ref{tab: nonGeVdet}. We also calculated the spectral points and upper limits using four logarithmically spaced bins across the energy range.

\subsection{GeV detected SFGs and the $L_{\gamma}$ -- SFR correlation}

\label{sec: GeVDet}

After over fifteen years of survey in the 50\,MeV$-$1\,TeV energy range, \textit{Fermi}-LAT has detected more than 7000 $\gamma$-ray sources. Their properties are accessible in the \textit{Fermi}-LAT Fourth Source Catalog Data Release 4 \citep[4FGL-DR4, gll$\_$psc$\_$v33.fit,][]{4fgl}. Only a small fraction of these sources have been associated with galaxies with active star formation, either normal galaxies, SBGs or Seyfert galaxies. We only consider galaxies that are included in the 4FGL-DR4. We do not include other possible sources that may show evidence of detection in the literature but do not meet the selection criteria discussed in section~\ref{sec: nonGeVDet}.\footnote{Of particular note is the ${\sim}\,4\sigma$ evidence of signal from the galaxy NGC\,1365 presented by \cite{Ambrosone2024}. The latter SFG is below the selection threshold for each of the three galaxy samples considered in section~\ref{sec: nonGeVDet}. Confirmation of its $\gamma$-ray emission may suggest contamination by the Seyfert nucleus, which is particularly prominent in X-rays.} The 4FGL-DR4 provides two types of association (ASSOC1/2) for each source. The first one corresponds to the name of the identified or likely associated source and the second one to a lower confidence association. For each association, the catalog also provides an object type designation \citep[CLASS1/2, see Table 6 in ][for further details]{4fgl}. As we aim to compile a comprehensive sample of galaxies with active star formation, we included sources with class designations 1 and 2 that align with normal galaxies  (``GAL''), starburst galaxies (``SBG''), or Seyfert galaxies (``SEY''). We found 29 $\gamma$-ray sources in the 4FGL-DR4 with at least one of these three classes. 

\begin{table*}
\caption{\label{tab: GeVdet} Sample of SFGs included in the 4FGL that satisfy the selection criteria defined in Sec. \ref{subsec: GeVDet_sel_crit}.}

\centering
\begin{tabular}{llccccc}
\hline\hline
Name 	&	 4FGL Name 	&	 $d$ 	&	 R.A. 	&	 Dec	&	 $\dot{M}_*$ 	&	 $\mathcal{F}_\mathrm{2\,GeV}$  \\    
 	&	 	&	 $\mathrm{Mpc}$ 	&	 deg 	&	 deg	&	 $M_{\odot}\,\mathrm{yr^{-1}}$ 	&	 $10^{-13}\,\mathrm{erg\,s^{-1}\,cm^{-2}}$ 	\\
\hline													
LMC 	&	 J0519.9-6845e    	&	 $0.049 \pm 0.002$   	&	 $\phantom{0}80.00$ 	&	 $-68.75$	&	 $0.192 \pm 0.022$   	&	 $240.0 \pm 5.9\phantom{00}$ \\    
SMC 	&	 J0058.0-7245e    	&	 $0.064 \pm 0.001$   	&	 $\phantom{0}14.50$ 	&	 $-72.75$	&	 $0.030 \pm 0.003$   	&	 $50.3 \pm 2.5\phantom{0}$  \\    
M31 	&	 J0043.2+4114     	&	 $0.74 \pm 0.02$   	&	 $\phantom{0}10.82$ 	&	 $\phantom{-}41.24$	&	 $0.245 \pm 0.023$  	&	 $3.91 \pm 0.55$ \\    
Circinus 	&	 J1413.1-6519 	&	 $2.41\pm 0.26$ 	&	 $213.29$ 	&	 $-65.33$	&	 $0.68 \pm 0.13 $                   	&	 $\phantom{0}8.3 \pm 1.2\phantom{0}$  \\    
NGC\,2403 	&	 J0737.4+6535 	&	 $3.13 \pm 0.06$ 	&	 $114.37$ 	&	 $\phantom{-}65.59$	&	 $0.355 \pm 0.012$  	&	 $1.79 \pm 0.39$   \\    
M82 	&	 J0955.7+6940     	&	 $3.53 \pm 0.03 $  	&	 $148.95$ 	&	 $\phantom{-}69.67$	&	 $10.41 \pm 0.19\phantom{0} $   	&	 $17.00 \pm 0.98\phantom{0}$  \\    
NGC\,253 	&	 J0047.5-2517 	&	 $3.61 \pm 0.03$  	&	 $\phantom{0}11.90$ 	&	 $-25.29$	&	 $\phantom{0}5.19 \pm 0.09\phantom{0}$ 	&	 $14.0 \pm 1.0\phantom{3}$  \\    
NGC\,4945 	&	 J1305.4-4928 	&	 $4.06 \pm 0.77$ 	&	 $196.36$ 	&	 $-49.47$	&	 $1.46 \pm 0.44 $  	&	 $16.50 \pm 0.90\phantom{0}$  \\    
NGC1068 	&	 J0242.6-0000 	&	 $13.4 \pm 1.8\phantom{0}$ 	&	 $\phantom{0}40.67$ 	&	 $\phantom{0}{-}0.01$	&	 $40 \pm 10$ 	&	 $9.12 \pm 0.68$  \\    
NGC\,2146 	&	 J0618.1+7819 	&	 $17.2 \pm 3.2\phantom{0}$ 	&	 $\phantom{0}94.53$ 	&	 $\phantom{-}78.33$	&	 $14.0\pm 5.1\phantom{0} $ 	&	 $2.49 \pm 0.38$  \\    
NGC\,3424 	&	 J1051.6+3253 	&	 $18.8 \pm 3.7 \phantom{0}$ 	&	 $162.91$ 	&	 $\phantom{-}32.89$	&	 $0.86 \pm 0.32$    	&	 $1.32 \pm 0.35$  \\    
NGC\,7059 	&	 J2127.6-5959 	&	 $24.7 \pm   4.6\phantom{0} $ 	&	 $321.84$ 	&	 $-60.01$	&	 $0.67 \pm 0.09$       	&	 $1.30 \pm 0.34$  \\    
Arp\,299 	&	 J1128.2+5831 	&	 $46.8\pm 3.3\phantom{0}$ 	&	 $172.07$ 	&	 $\phantom{-}58.52$	&	 $\phantom{0.}97 \pm 14\phantom{.0}$ 	&	 $2.00 \pm 0.37$  \\    
Arp\,220 	&	 J1534.7+2331 	&	 $80.9 \pm 5.7\phantom{0} $ 	&	 $233.70$ 	&	 $\phantom{-}23.53$	&	 $214\phantom{.} \pm 32\phantom{.0}$ 	&	 $2.99 \pm 0.53$  \\    
\hline\hline
\end{tabular}

\tablefoot{The first two columns provide the common name of the galaxy and the name of the associated $\gamma$-ray source in 4FGL-DR4, $d$ is the luminosity distance, R.A. and Dec are the right ascension and declination in ICRS reference system of the center of the associated 4FGL source. $\dot{M}_*$ is the SFR. $\mathcal{F}_\mathrm{2GeV}$ is the energy flux at 2\,GeV calculated using the best-fit power-law parameters from the 4FGL-DR4.}
\end{table*}

\label{subsec: GeVDet_sel_crit}
Associations of \textit{Fermi}-LAT sources are established when there is a statistically-significant positional coincidence with a counterpart at another wavelength that is an a priori candidate for $\gamma$-ray emission \citep{1FGL}. The automated source association carried by the \textit{Fermi}-LAT Collaboration relies on curated catalogs containing potential counterparts. The classification of the same source may vary depending on the wavelength and selection criteria used in catalog generation. Because of the limited accuracy of source localization, additional data, such as temporal variability and/or spectral characteristics, are taken into account to strengthen the associations. This proves to work well for AGN sources but is  trickier for quieter sources like non-jetted AGNs or SFGs.

To address this issue, we explored each candidate association.
Stellar light is particularly prominent in the broad-band spectrum of SFGs \citep{Kennicutt2012}. Such light reaches us directly from the emitting stars or is reprocessed in the infrared by dust present in the surrounding interstellar medium (ISM). Thus, SFGs show a continuum spectrum with two prominent peaks in the optical and infrared bands, respectively \citep{Cunha2008}. This shape serves as a valuable indicator of active star formation within a galaxy. The presence of a star-forming component in the optical and infrared spectrum of the \textit{Fermi}-LAT candidates was thus evaluated using the  ``SSDC SED Builder''.\footnote{https://tools.ssdc.asi.it/SED/} 

Out of the 29 \textit{Fermi}-LAT candidates, we have excluded 15 galaxies that do not show clear evidence of a star-forming component at low energies. Among these are NGC\,5380 and IC\,678, which are classified as normal galaxies in 4FGL-DR4 but with spectra that are better represented by that of a giant elliptical galaxy (see Table~\ref{tab: excl}  in Appendix~\ref{app: complement} for further details). Our study includes the $\gamma$-ray source 4FGL\,J0737.4+6535, classified as a SBG and labeled as the infrared source WISEA\,J073707.21+653623.0 in the 4FGL-DR4 catalog. This infrared source was associated with a blazar by \citet{Bruzewski2023}. The 4FGL source is also coincident with the disk of the SBG NGC\,2403 detected by \citet{Ajello2020}. As a starting point, we assume that 4FGL\,J0737.4+6535 is associated with the galaxy NGC\,2403.

\label{subsec: correl}

The correlation between SFR and the $\gamma$-ray luminosity ($L_{\gamma}$) in different energy ranges has been explored by several authors \citep{Lacki2011,Ackermann2012, Ajello2020,K20,Roth2021}. As it is commonly accepted that the emission in the GeV range comes from proton-proton interactions, the spectrum in this energy range is expected to follow a power law with an index close to that of CRs injected in the star-forming environment. The $\gamma$-ray luminosity is expected to increase with increasing SFRs, so that the normalization of the $\gamma$-ray spectrum depends mostly on the SFR value and on the distance to the source. It should be noted that the $\gamma$-ray slope could differ from the slope of the injected CRs depending on energy due, e.g., to CR escape by diffusion or to absorption. Thus, the correlation between $\gamma$-ray luminosity and SFR provides insight into the mechanisms and physical properties that govern the transport and dissipation of CRs within SFGs \citep{K22}.

Diffusion may become dominant at sufficiently high energies (depending on the turbulence level and the size of the emission region), leading to a softening of the spectrum and inducing a transition regime in which protons can escape the galaxy before they interact and radiate. Particles may also escape due to the winds of SFGs \citep{Veilleux2005}, which can advect protons across the entire energy range, diminishing the overall normalization of the $\gamma$-ray spectrum without altering the slope \citep{Peretti2019, K22}. This effect has been carefully studied in NGC\,253 and its impact, which depends on wind speed, seems to be minor \citep{HESS2018}; nevertheless, it could be more pronounced for galaxies with stronger winds, such as M82 \citep{Strickland2009}.

Since SFGs can host a strong infrared field, primarily emitted by hot dust, $\gamma$$\gamma$ absorption may occur between these target photons and $\gamma$-ray photons. The relevance of this absorption mechanism starts to become significant at energies close to ${\sim}\,10\,$TeV \citep{Peretti2020} and increases at higher energies. It can be even more pronounced in galaxies with intense radiation fields, such as ULIRGs.
At energies beyond 20\,TeV for nearby galaxies (the most distant galaxy in our sample is Arp\, 220 at $z \sim 0.019$), the extragalactic background light (EBL) and cosmic microwave background become the dominant target photon fields for $\gamma$$\gamma$ absorption \citep[e.g.][for a review]{2022Galax..10...39B}.

Here we revisit the $L_{\gamma}$ -- SFR correlation presented in \cite{Ackermann2012}. Instead of using the luminosity integrated over the energy range of \textit{Fermi}-LAT, we use the luminosity at a reference energy. We set the reference energy at 2\,GeV as, for the $\gamma$-ray sources in our sample, it is close to the pivot energy of the best-fit power law, where the uncertainty on the $\gamma$-ray flux is the smallest. We obtained the 2\,GeV energy flux and associated uncertainty using the best-fit power-law parameters from the 4FGL-DR4. For NGC\,2146, Arp\,220, and Arp\,299, luminosity distances not listed in Cosmicflows-4 \citep{Tully2023} were adopted from \citet{K20}. Except for NGC\,7059, SFRs were recalculated following \citet{K20} and references therein, using updated distances.
For NGC\,7059 we took the IR$_{25 \mu \rm{m}}$ flux from \citet{rice1988} and the FUV flux from \citet{Bouquin2018} to estimate the SFR. We report these values in Table~\ref{tab: GeVdet}.

\begin{figure}
  \centering
  \includegraphics[width=0.45 \textwidth]{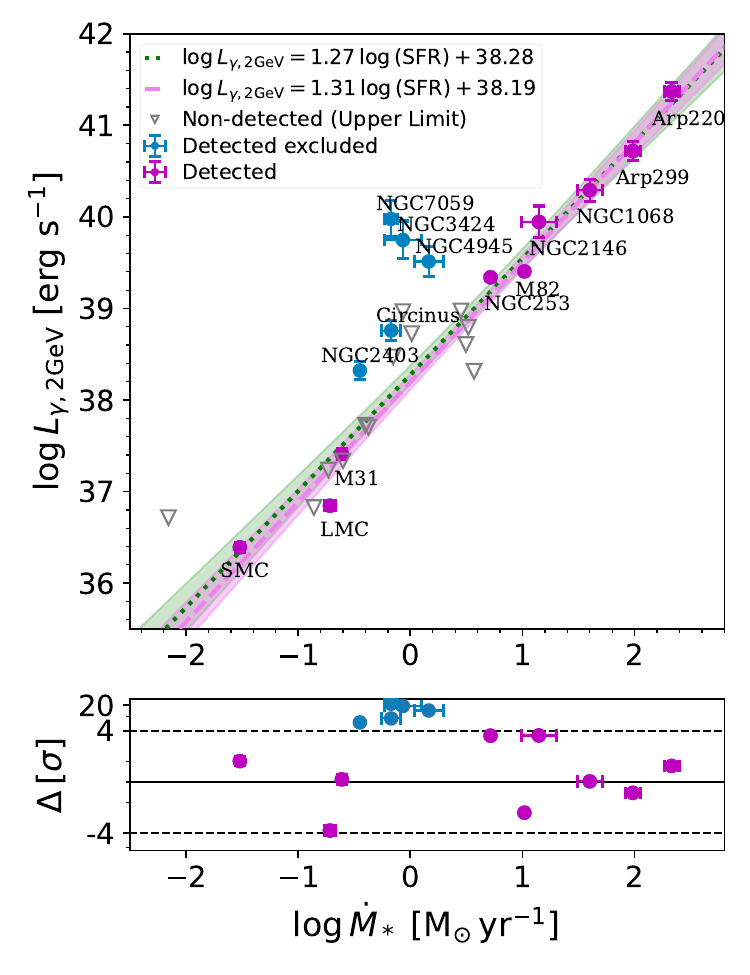}
\caption{\textit{Upper panel:} $L_{\gamma,\, \text{2\,GeV}}$ -- SFR observed correlation. 
The dotted green line corresponds to the best-fit model for all the data points of detected SFGs and the shaded region represents the 68$\%$ confidence-level band.
The dashed pink line and associated shaded region shows the best-fit model excluding blue markers. The upper limits obtained in section~\ref{sec: nonGeVDet}, shown as downward pointing triangles, are not taken into account in the fit. \textit{Lower panel:} residuals to the fit excluding blue markers normalized to their standard deviation. The black dashed lines correspond to 4 standard deviations.}
  \label{fig: corr}
\end{figure}

In the upper panel of Fig.~\ref{fig: corr}, we show the $\gamma$-ray luminosity at 2\,GeV of SFGs, $L_{\gamma,\, \rm{2\,GeV}}$, as a function of their SFR, $\dot{M}_*$. We fit the data in Table~\ref{tab: GeVdet} with the orthogonal regression method to include the errors in both $L_{\gamma}$ and SFR, using the ODRPACK  software.\footnote{More details about the fitting procedure can be found in the SciPy documentation \url{https://docs.scipy.org/doc/external/odrpack_guide.pdf}} We also show in Fig.~\ref{fig: corr} the upper limits obtained in section~\ref{sec: nonGeVDet}, which we do not include in the fit.\footnote{Properly including undetected sources would require a Bayesian analysis, taking into account the likelihood profile of the $\gamma$-ray flux for each of the sources. \citet{Ambrosone2024} estimate that the normalization of the relation could be impacted by a factor of two if non-detected galaxies are accounted for.}
A power-law fit, $L_{\gamma,\, \rm{2\,GeV}}$ = C $\dot{M}_*^m$, where $L_{\gamma, \rm{2\,GeV}}$ is in $\rm{erg}\, \rm{s}^{-1}$ and $\dot{M}_*$ is in $M_{\odot}\, \rm{yr}^{-1}$, to all the data (pink and blue dots in Fig.~\ref{fig: corr}) yields a slope value of $m = 1.27 \pm 0.10$, and a constant $\log C = 38.28 \pm 0.10$, with a residual variance of $25$. The residual variance is calculated as the average squared orthogonal distance from the data points to the fitted model, accounting for errors along both axes, and normalized by the associated number of degrees of freedom. We show this fit as a green dotted line in Fig.~\ref{fig: corr}. Most galaxies follow a clear trend, except NGC\,3424, NGC\,4945, Circinus, NGC\,2403 and NGC\,7059.

If we exclude these five galaxies from the sample, the correlation is tighter, with $m = 1.31 \pm 0.08$, $\log C = 38.19 \pm 0.08$ and a smaller residual variance of $14$. We show the corresponding fit as a pink dashed line in Fig.~\ref{fig: corr}. The lower panel of Fig.~\ref{fig: corr} shows the pulls, i.e.\ residuals for this fit normalized to the standard deviation. The figure show that NGC\,3424, NGC\,4945, Circinus, NGC\,2403 and NGC\,7059 stand out by more than $4\sigma$ from the best fit (at  $19$, $14$, $8.7$, $6.8$ and $22\sigma$ respectively).

The deviation of these galaxies from the correlation followed by the other SFGs could potentially stem from an additional contribution associated with an AGN as was discussed by \citet{Gavazzi2011} for NGC\,3424 and NGC\,2403, \citet{Lenc2009} for NGC\,4945, and \citet{Prieto2004} for Circinus. Such a deviation could also be attributed to a misestimation of their SFR \citep[see discussion in][]{K20}{}. The emission coming from NGC\,2403 could further be contaminated by a supernova explosion as discussed by \cite{Xi2020}, or it could  just be misassociated and, as  suggested by \citet{Bruzewski2023}, be a blazar. Finally, the $\gamma$-ray source 4FGL\,J2127.6-5959, associated with the galaxy NGC\,7059 in the 4FGL-DR4, is the one that deviates the most from the correlation. Interestingly, the $\gamma$-ray source shows a power-law index of $1.8 \pm 0.1$, which is marginally outside the expected and observed values for the rest of the sample.

\section{Candidate SFGs in the TeV energy range}
\label{sec:TeVcandidates}

\subsection{TeV Gamma-Ray spectra and Sensitivity}

The best-fit model for the correlation discussed in Section~\ref{subsec: correl} provides an estimate of the $\gamma$-ray luminosity of a SFG at 2\,GeV  based on its SFR. Using the correlation and the expected index value of 2.2, we developed a simple empirical model to estimate the flux of SFGs. We assume a power-law spectral energy distribution (SED) with a constant slope of $\alpha = 2.2$ from GeV to TeV energies and a $\gamma$-ray luminosity provided by the best-fit model to the pink markers in Fig.~\ref{fig: corr}. It should be noted that the $\gamma$-ray emission from some of these sources could deviate from a power law. This is the case, e.g., for M31, NGC\,253, and M82, the $\gamma$-ray spectra of which are better modeled by a log-parabola in the 4FGL. However, the curvature significance is not substantial (approximately 3$\sigma$). Therefore, the power-law hypothesis remains appropriate. We use the EBL model developed by \citet{Dominguez2011} to account for $\gamma$$\gamma$ absorption on Mpc scales in the intergalactic medium. The $\gamma$-ray energy spectrum is thus estimated as follows:
\begin{equation}
\mathcal{F} [\mathrm{erg} \, \mathrm{s}^{-1} \mathrm{cm}^{-2}] = \frac{L_{2 \, \mathrm{GeV}}(\mathrm{SFR})}{4 \pi d^2} \times \left( \frac{E}{2 \, \mathrm{GeV}} \right)^{2-\alpha} \mathrm{e}^{-\tau(E, z)},
\label{eq}
\end{equation}
where $\tau(E, z)$ is the optical depth from the model of \citet{Dominguez2011}. To compute $z$ from the luminosity distance $d$, we consider a Hubble constant of $H_0 = 67.4\, \rm{km}\, \rm{s}^{-1} \,\rm{Mpc}^{-1}$ and a matter density of $\Omega_{\rm{m}} = 0.315$ \citep{plank2020}. 

We also evaluate an even simpler model for the sources detected at GeV energies, namely the extrapolation of their spectrum measured by \textit{Fermi}-LAT with EBL attenuation at the highest energies. For SFGs not included in the 4FGL, for which we do not have an observed spectrum, we provide an upper bound on the energy flux at 2\,GeV, $\mathcal{F}_\mathrm{2\,GeV}^{\mathrm{UL}}$, as estimated from the upper limits on the integrated energy flux, $\mathcal{F}^\mathrm{UL}_\mathrm{0.1-100\,GeV}$ derived from \textit{Fermi}-LAT data, assuming a photon index of 2.2. 

The relation in Eq.~\ref{eq} or other extrapolations to the highest energies are strictly valid in the energy range where internal absorption is subdominant. As discussed in the previous subsection, internal-absorption effects are expected to occur at energies of the order of tens of TeV and to depend on the intensity of the galaxy infrared field.  We do not consider the modeling of internal absorption within the galaxy, as we aim to maintain a simple, geometry-independent model in our study; however, we discuss the implications this may have when drawing conclusions. Spectral effects induced by CR escape could be noticeable at energies close to 10\,TeV but they would not completely change the spectral slope, except in the extreme case of fast diffusion as in Milky-Way like galaxies \citep{Strong2011, Do2021}. Therefore, the results from our simple model should be considered optimistic beyond 10\,TeV.


\label{sec: TeVsens}

As mentioned in the introduction, the best new-generation telescopes for the detection and study of SFGs at TeV energies are expected to be the CTAO, LHAASO and SWGO. The CTAO, which is currently under construction, will be situated at the Paranal Observatory in Chile and at the Instituto de Astrofísica de Canarias, in La Palma, Spain \citep{CTABOOK2019}. Using \emph{Gammapy}\footnote{https://docs.gammapy.org/1.1/} \citep{Donath2023}, we estimated the point-source sensitivity for 50\,h of CTAO observation assuming 20 logarithmic energy bins across the energy range $0.03-30$\,TeV. The significance in each bin is computed for a 1D analysis (ON-OFF regions) assuming a fixed source offset of 0.5\,degree from the pointing position. As we investigate which galaxies may approach the detection threshold of the CTAO, we impose a minimal number of 5 signal counts per bin and a minimal significance of $3\sigma$ per bin, for a background  fraction of 0.1. We have verified with dedicated simulations that comparing the spectra of SFGs with these differential sensitivity curves provides a good estimate of detectability: for a photon index of 2.2, a spectrum above the sensitivity curve corresponds to an expected significance of the source, combined over all energy bins, above $5\sigma$.
We computed the sensitivity for an azimuth-averaged pointing using the latest version of the CTAO instrument response functions (IRFs): prod5 v0.1.\footnote{https://zenodo.org/record/5499840} The IRFs of the Northern and Southern arrays (CTAO-N and CTAO-S, respectively) are calculated for three different zenith angles (zen): 20, 40 and 60\,degrees. We determined the number of hours of darkness for each array using the TeVCat Object Visibility Tool,\footnote{http://tevcat.uchicago.edu/CustomVis.pl} and kept the IRF with the smallest zenith angle for which more than 50 hours of observation are available over a year for each galaxy. We discarded the low energy points on the sensitivity curve where the IRF is not well defined (implied energy threshold: CTAO-S, zen = 40 deg, $E > 0.05$\,TeV; CTAO-S, zen = 60 deg, $E > 0.2$\,TeV; CTAO-N, zen = 60 deg, $E > 0.07$\,TeV).

LHAASO is a multi-component facility located in Daocheng, China, that serves as a $\gamma$-ray telescope operating in the energy range from 0.1\,TeV to 1\,PeV. Due to its geographical location, most of the LHAASO field of view is in the Northern hemisphere, with access to sources of declination larger than $-30$\,degrees \citep{LHAASO2019, Yang2019}. Therefore, we compared the expected $\gamma$-ray flux of SFGs with the LHAASO sensitivity \citep[with 20'' photomultiplier tubes,][]{LHAASO_sens2021} in cases of galaxies with such declinations. SWGO \citep{Albert2019} is a forthcoming $\gamma$-ray observatory in South America, featuring ground-level particle detection technology with a wide field of view spanning several steradians. Focused on energies from hundreds of GeV to the PeV scale, SWGO will primarily use water Cherenkov detectors. The experiment is planned for construction between 10 and 30 degrees South latitude to ensure a clear view of the Galactic Center. We then compare the SWGO optimistic sensitivity to sources with negative declination. This sensitivity is estimated for a steady point source at a zenith angle of 20\,deg and under the assumption that it can be observed for 6 hours per day \citep{Albert2019}. The differential sensitivity curves of LHAASO and SWGO typically allow for 5 to 10 bins per decade in energy. To claim a detection, the number of excess events (over the background) is typically required to be above 10 events, and the significance threshold for detection is generally set at $5\sigma$. 

As the calculation of the CTAO sensitivity to extended sources is more complex than for point sources and falls beyond the scope of this study, five galaxies in the Local Group deserve a specific discussion due to their proximity, namely LMC, SMC, NGC\,6822, M31 and M33. Assuming that most of the emission is located near the central molecular zone of radius $R < 1 \, \text{kpc}$, the star-forming nucleus would appear ``point-like'' if it were viewed within a cone of half-opening angle $\theta \approx R/d < \theta_{68}(0.2 \, \text{TeV})$, where $\theta_{68}(0.2 \, \text{TeV}) \approx 6'$ is the 68\% containment radius at 0.2\,TeV for the CTAO.\footnote{see \url{https://www.ctao.org/for-scientists/performance/}} Point-like emission would then be expected for SFGs at $d > R/\theta_{68} \approx 600 \, \text{kpc} \times (R/1 \, \text{kpc}) \times (\theta_{68}/6')^{-1}$. With distances of 740\,kpc and 850\,kpc, M31 and M33 could potentially be detected as point-like sources and will be considered in this work for comparison with the sensitivity of new-generation $\gamma$-ray observatories. On the contrary, the central molecular zones of the SMC and LMC, with distances of 62\,kpc and 50\,kpc, respectively, could appear as extended sources and require dedicated studies \citep[e.g.][]{LMCCTA2023}. Therefore, we will omit the direct comparison of the SMC and LMC spectra with the point-source sensitivity of $\gamma$-ray observatories. At a distance of 460\,kpc, NGC\,6822 could show a mild extension. We assume that the source is punctual and will show that even in this optimistic scenario the source is difficult to detect.

\subsection{Results: detectability in the TeV energy range}

By comparing the CTAO point-source sensitivity with the extrapolated best-fit power-law spectrum from 4FGL-DR4, or the upper limits for SFGs not included in 4FGL, we can identify the most promising candidate point sources at TeV energies.

\begin{figure*}[ht!]
  \centering
    \includegraphics[width=0.75\textwidth]{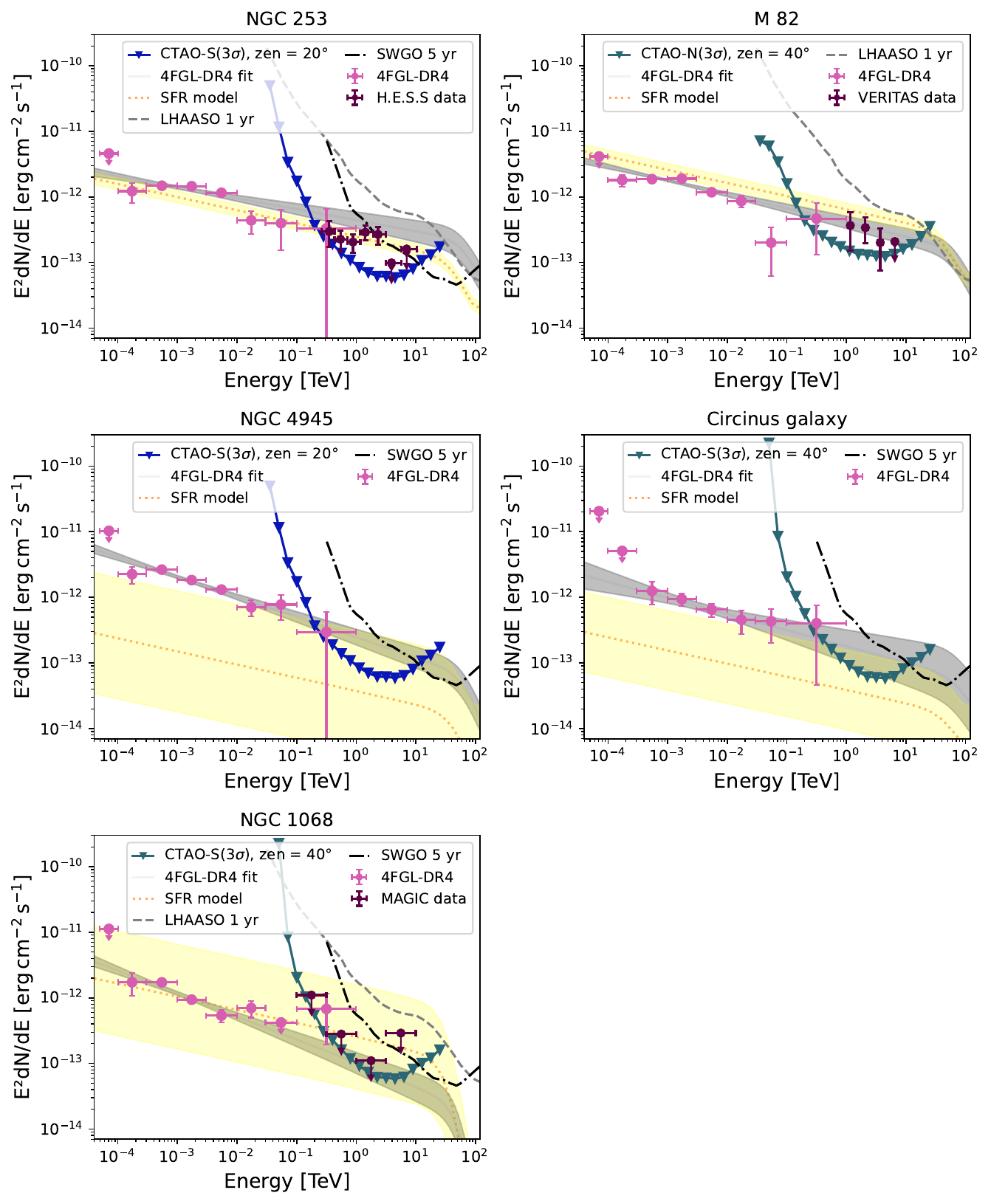}
\caption{SEDs of the top candidates selected from the GeV-detected sample of SFGs. In each panel, pink dots denote spectral points from \textit{Fermi}-LAT, the gray line and gray shaded region represent the best power-law fit to the $\gamma$-ray data, as provided in the 4FGL-DR4 catalog, including absorption on the EBL at multi-TeV energies. The orange dotted line and associated shaded region depict the empirical model scaled to the SFR of the galaxy along with its associated uncertainty. The line defined by the triangles indicates the $3\sigma$-level point-source sensitivity of the CTAO in 50\,h. Gray dashed line and black dotted-dashed line show the LHAASO and SWGO sensitivity respectively. VHE data and upper limits for NGC\,253, M82 and NGC\,1068 were taken form \citet{HESS2018, VERITAS2009} and \citet{MAGIC2019}, respectively.}
  \label{fig: SED_best_det}
\end{figure*}

\begin{figure}[ht!]
\centering
  \includegraphics[width=0.42 \textwidth]{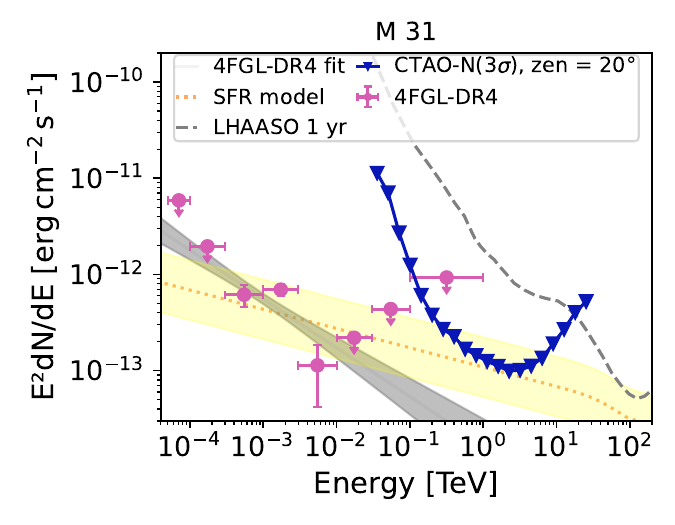}\\
  \includegraphics[ width=0.42 \textwidth]{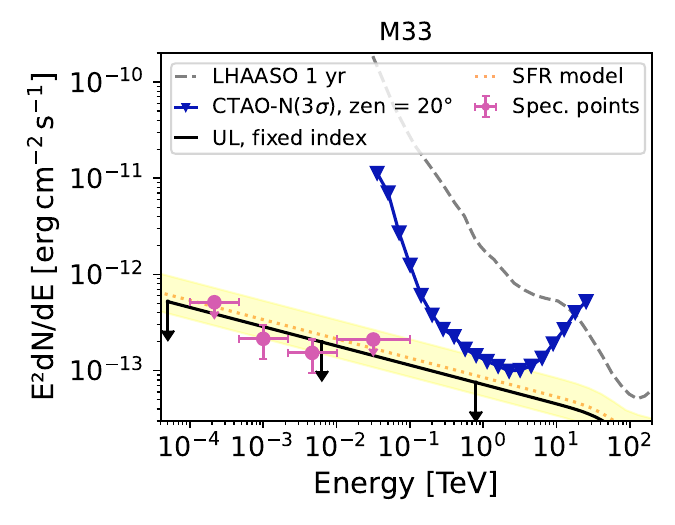}\\
  \includegraphics[width=0.42\textwidth]{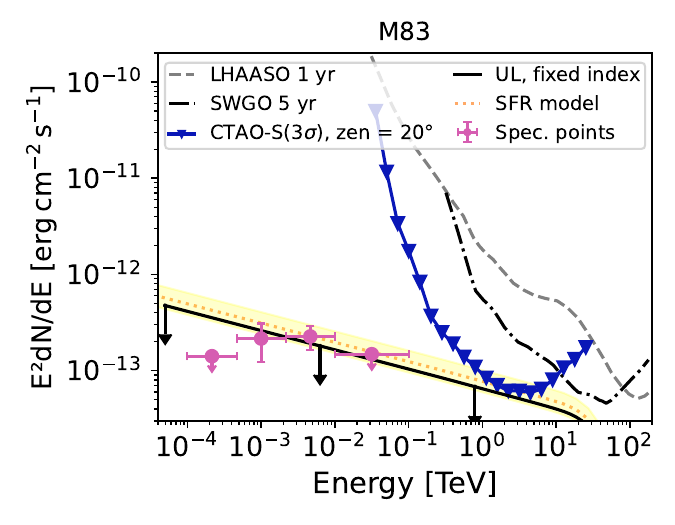}
\caption{SEDs of interesting candidates detected at GeV energies, namely M31 (\textit{upper panel}), or on the verge of GeV detection, namely M33 (\textit{mid panel}) and M83 (\textit{lower panel}). The line and color code match that of Fig.~\ref{fig: SED_best_det}. The black solid line with arrows represents the upper bound on the expected $\gamma$-ray emission of each galaxy including the absorption on the EBL at multi-TeV energies.}
  \label{fig: SED_best_non_det}
\end{figure}

\begin{figure*}[ht!]
  \centering
\includegraphics[width=0.75\textwidth]{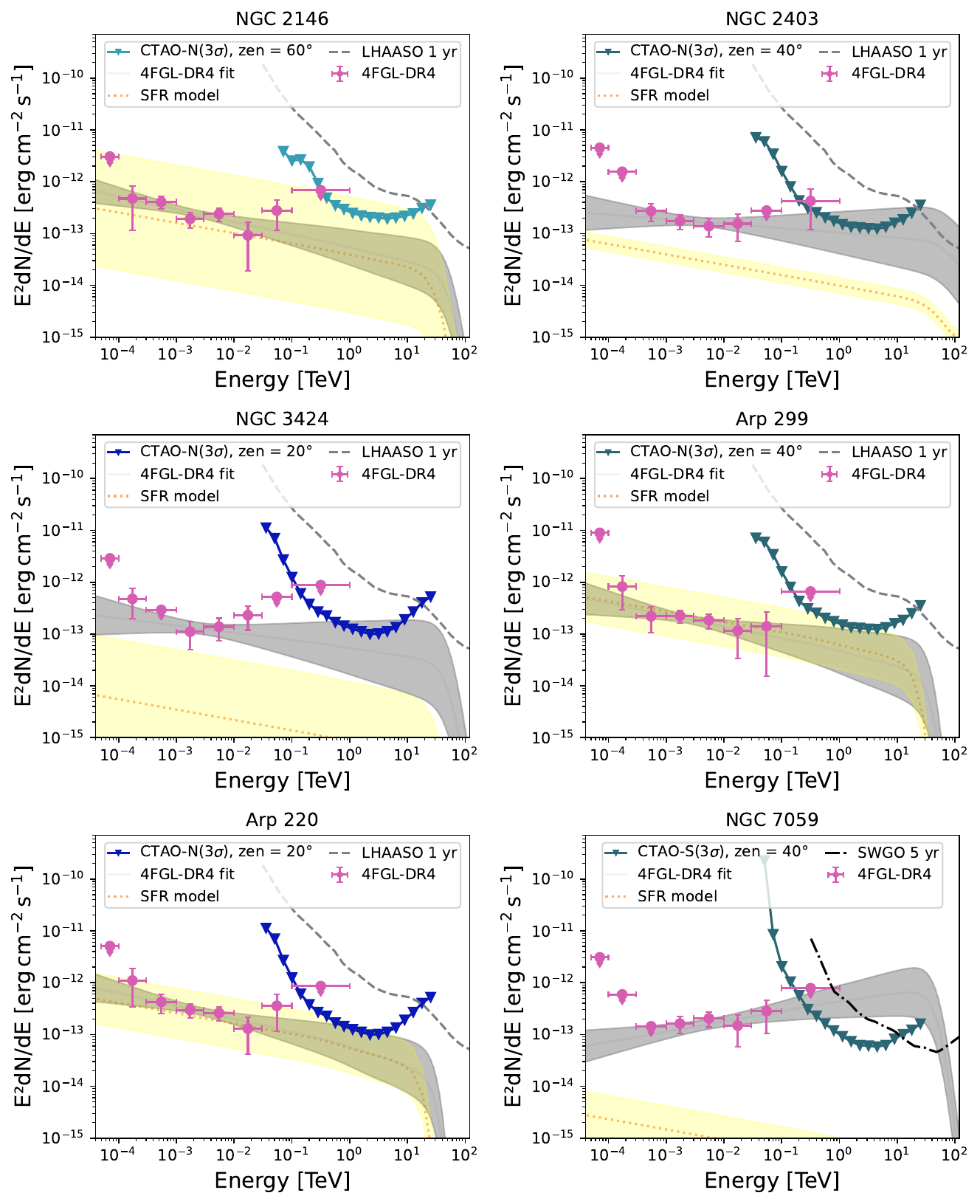}
\caption{SEDs of possibly interesting candidates from the GeV detected sample. The line and color code match that of Fig.~\ref{fig: SED_best_det}.}
  \label{fig: SED_nonbest_det}
\end{figure*}

In Fig.~\ref{fig: SED_best_det}, we show the SED of SFGs detected at GeV energies that are the best candidates for detection in the TeV range by the CTAO, LHAASO and SWGO. The extrapolated $\gamma$-ray spectra are above the CTAO point-source sensitivity for NGC\,253, M82, NGC\,1068, NGC\,4945 and Circinus. Figure~\ref{fig: SED_best_non_det} shows the interesting candidates for which GeV $\gamma$-ray observations could still allow for TeV detection. The next best candidates are shown in Fig.~\ref{fig: SED_nonbest_det}. The SEDs of the other SFGs that are not selected as strong candidates for detection at TeV energies are provided in Appendix~\ref{app: complement}.

Among the SEDs of GeV-detected SFGs listed in Table~\ref{tab: GeVdet}, the empirical model scaled with SFR is consistent with the extrapolated $\gamma$-ray spectrum from \textit{Fermi}-LAT for all galaxies but NGC\,4945, NGC\,3424, NGC\,2403 and M31. Such a discrepancy was expected for the first three galaxies as their $\gamma$-ray luminosity expected from the correlation in Fig.~\ref{fig: corr} differs by more than $4\sigma$ from the observation. Regarding M31, the discrepancy is due to the fact that its observed photon index ($\alpha \approx 2.5$) is softer than that assumed in the empirical model scaled to SFR. 
Normal SFGs (with low SFRs) exhibit less dense environments, potentially reducing the effective density of ambient gas encountered by CR protons and facilitating their escape \citep[e.g.][]{K20}. This context can enhance the relative contributions of leptonic emission and unresolved point-source populations to the integrated $\gamma$-ray emission \citep{Do2021}. Consequently, the final spectrum observed by \textit{Fermi}-LAT may deviate from the slope predicted by our simple model.

For SFGs not detected in the GeV energy range, the upper-bound spectrum is shown as a black line with downward arrows in Fig.~\ref{fig: SED_best_non_det} and as colored lines in the Appendix (Fig.~\ref{fig: SED_other}). As shown in Fig.~\ref{fig: SED_best_non_det}, the flux of M83 and M33 inferred from the scaling of the $\gamma$-ray luminosity with SFR is at the threshold of the sensitivity of the CTAO, while the upper limit inferred from GeV $\gamma$-ray observations could still allow for TeV detection provided a sufficient exposure. We do not include NGC\,7059 in the best-candidate sample although the extrapolation of the spectrum of the associated  \textit{Fermi}-LAT source is above the CTAO sensitivity curve. As illustrated in Fig.~\ref{fig: SED_nonbest_det}, the empirical model scaled with SFR differs strongly from the extrapolated $\gamma$-ray spectrum for NGC\,7059, a discrepancy which is amplified at the highest energies by the hard spectrum of the $\gamma$-ray source. This discrepancy suggests that the dominant $\gamma$-ray emission does not originate from the star formation in the disk of this galaxy \citep{2019ApJ...887...18K}. 

We discuss the sources in more detail on a case-by-case basis in the next section. A summary of the prospects for TeV detection is also given in Table~\ref{tab: summary}.

\subsection{Case-by-case discussion}

\label{sec: disc}

\paragraph{NGC\,253 and M82 (see Fig.~\ref{fig: SED_best_det}):}
In the TeV energy range, NGC\,253 and M82 show empirical model spectra and extrapolated spectra from \textit{Fermi}-LAT data that are above the 50h sensitivity of the CTAO over almost one and a half orders of magnitude in energy. These two SBGs have already undergone dedicated investigations by the CTAO Consortium \citep{CTABOOK2019}. Furthermore, the extrapolated model for NGC\,253 and both models for M82 are above the sensitivity of LHAASO beyond 30\,TeV.
The extrapolated and empirical models for NGC\,253 are also above the  sensitivity of SWGO. For these two observatories, the detectability of NGC\,253 and M82 strongly depends on the level of internal absorption within the sources \citep[see][for M82]{Peretti2020}.

\paragraph{NGC\,1068, NGC\,4945 and Circinus (see Fig.~\ref{fig: SED_best_det}):}
Albeit with a narrower margin, the extrapolated spectra of NGC\,1068, NGC\,4945 and Circinus are above the CTAO sensitivity curve.
The extrapolated models are consistent with the empirical models for NGC\,1068\footnote{We checked that using the commonly quoted distance of 10.1\,Mpc for NGC\,1068 \citep{2011A&A...532A.104N} has a negligible impact on the empirical model in Fig.~\ref{fig: SED_best_det}.} and Circinus. This is not the case for NGC\,4945: its empirical model falls below from the extrapolation from GeV observations. As mentioned above, this is not surprising as this galaxy is an outlier from the best-fit correlation in Fig.~\ref{fig: corr}. A higher SFR or an additional non-thermal contribution from either an AGN or a super wind \citep{Perez2011} could bring the two models in agreement. 
Previous studies excluded NGC\,4945 and NGC\,1068 from consideration due to the potential presence of AGNs \citep{Shimono2021}. Without clear evidence that the GeV $\gamma$-ray emission originates from the AGN, we studied these galaxies and highlight NGC\,1068 and NGC\,4945 as potential targets for CTAO observations. 
The extrapolated spectra of NGC\,4945 and Circinus galaxy are above the sensitivity of SWGO for energies greater than 10\,TeV. As for NGC 253 and M82, observations above 10\,TeV, together with those at TeV energies, will lead to constraints on the amount of internal absorption.

\paragraph{M83 and M33 (see Fig.~\ref{fig: SED_best_non_det}):}
These galaxies are marginally detected by \textit{Fermi}-LAT with photon indices of $2.1 \pm 0.2$ (TS\,=\,18) and $2.4 \pm 0.2$ (TS\,=\,22), respectively. Previous excesses near the position of M83 have been reported by \citet{Xing2023} and \citet{Ambrosone2024}, with TS values of 14 between $0.1-500$\,GeV and 15 between $1-1000$\,GeV, respectively. 
\citet{Xi2020A} and \citet{Ajello2020} have claimed the detection of M33 \citep[photon index of $2.41 \pm 0.16$ in ][]{Ajello2020}. \citet{Xi2020A} argues that the $\gamma$-ray excess from M33 does not originate from the center but rather from a region in the northeast that is positionally coincident with a supergiant HII region. Nevertheless, this source remains absent from the 4FGL-DR4 catalog. Further data from \textit{Fermi}-LAT will enable a more accurate determination of the GeV spectrum of M33 and of its extrapolation to higher energies.

The empirical model and the upper bound on the extrapolated $\gamma$-ray emission are in line for M83 and M33. The spectral model for M83 and M33 are close to the sensitivity of the CTAO. In line with the predictions of \citet{Shimono2021}, we identify M33 and M83 as interesting candidates for observation in the TeV energy range.

\paragraph{M31 (see Fig.~\ref{fig: SED_best_non_det}):}

The extrapolated spectrum of M31 falls below the 50h sensitivity of the CTAO.
Despite being cataloged as a point-like source in the 4FGL catalog, M31 (aka 4FGL\,J0043.2+4114) has been suggested to exhibit marginal $\gamma$-ray extension by \citet{Abdo2010,Ackermann2017, Ajello2020}. Depending on the spatio-spectral model employed, the best-fit photon index varies considerably, from $2.1 \pm 0.3$ to $2.5 \pm 0.1$ \citep{Abdo2010, 4fgl}. \citet{McDaniel2019} proposed that M31 $\gamma$-ray emission is a mix of pionic and Compton scattering from primary and secondary electrons on ambient radiation fields. Alternatively, \citet{Yi2023} suggest it may arise from two different point sources: one at the galaxy center, likely from unresolved objects like millisecond pulsars \citep{Eckner2018}, and another 0.4 degrees away, with an unclear origin. Additionally, \citet{Persic2024} propose a more complex scenario, involving a combination of diffuse pionic, pulsar, and nuclear black-hole emissions in M31 core.
Thus, M31 remains an exceptional candidate for exploring the contribution of additional source populations and is also crucial for studying the $L_\gamma$ -- SFR correlation, as it is one of the few candidates with a low SFR.

\paragraph{NGC\,2146, NGC\,2403, NGC\,3424 (see Fig.~\ref{fig: SED_nonbest_det}):} 
The extrapolated spectrum is consistent with the empirical model only for NGC\,2146, while a tension is observed for NGC\,2403 and NGC\,3424, suggesting that an additional emission may be needed (see Sec.~\ref{subsec: correl}). The extrapolated spectra of these three sources are below the 50h sensitivity of the CTAO, although NGC\,2403 could be marginally detectable with the CTAO, and with LHAASO if not prevented by internal absorption. With updated \textit{Fermi}-LAT data, contrary to what was anticipated by \cite{LHAASO2019}, we do not find sufficient emission from NGC\,2146 at the highest energies for detectability by LHAASO in 1\,yr.

\paragraph{Arp\,299 and Arp\,220 (see Fig.~\ref{fig: SED_nonbest_det}):} 

The empirical model closely matches the \textit{Fermi}-LAT extrapolation for these galaxies. Both models are below the 50h CTAO sensitivity and the 1 year sensitivity of LHAASO. While Arp\,220 has been proposed as part of the key science project of the CTAO Consortium dedicated to star formation, other galaxies, such as Circinus, NGC\,1068, NGC\,4945 (and even NGC\,2403), emerge as more promising candidates for detection. The case of Arp\,299 bears similarities to that of Arp\,220.

Although they are not among the preferred candidates, it is crucial to keep an eye on these luminous infrared galaxies for further investigation as improved measurements of their GeV fluxes may grant a TeV flux close to the sensitivity  limit of the CTAO. Incorporating at least one of these objects in the observation program of the CTAO is pivotal for exploring a possible correlation between TeV emissions and SFR \citep{CTABOOK2019, Kornecki:2023Ia}, as at least one object per decade in SFR is necessary to comprehensively assess the influence of CRs on SFGs. This could involve extending observation time, for example, to 100 hours, and conducting a more exhaustive analysis of their $\gamma$-ray spectra, leveraging their multiwavelength emissions as constraints.

\paragraph{NGC\,7059 (see Fig.~\ref{fig: SED_nonbest_det}):} The source 4FGL\,J2127.6-5959 has been discussed by \cite{2022Univ....8..587F}, who questioned its association with the SFG NGC\,7059 due to inconsistent sky locations and counterparts observed at radio and X-ray wavelengths. As mentioned previously, NGC\,7059 not only shows an unexpected normalization with respect to the $L_{\gamma, \rm{2GeV}}$–SFR correlation but also exhibits a hard photon index more aligned with blazar emission \citep{2019ApJ...887...18K}. Nevertheless, the association of this source with NGC\,7059 cannot be ruled out, necessitating further studies to evaluate its true counterpart \citep{2022Univ....8..587F}. Assuming absorption on the EBL for a distance consistent with that of NGC\,7059, the model extrapolated from \textit{Fermi}-LAT data exceeds the sensitivity curves of both the CTAO and SWGO. Observing this source at VHE would be crucial to measure its cutoff energy, providing key insight into the association of 4FGL\,J2127.6-5959.

\paragraph{M106, M51, NGC\,1569 and NGC\,6822 (see Fig.~\ref{fig: SED_other}):}
These galaxies show one energy bin with TS $>$ 4 in in the GeV energy range. This could suggest a simple and expected background fluctuation or, alternatively, it might be a remote indication of an excess linked to the galaxy emission. The aforementioned scenario is more likely to apply to M106 and possibly M51 (TS $=$ 13 and 7, respectively), due to their high Galactic latitude. The excess associated with M106 is dominated by low-energy photons (best-fit photon index close to a value of 5) increasing the likelihood that these photons originate from nearby regions rather than being directly connected to emission from the galaxy. Although M51 was identified as an interesting candidate at GeV energies by \citet{RojasBravo2016}, the empirical model of its VHE flux and the extrapolated upper limit do not favor its observation in the TeV range.

The extrapolated upper limit of NGC\,6822 (TS $=$ 8) is close to the CTAO detection limit. This galaxy, as NGC\,1569 (TS $=$ 10), is predicted to have a low $\gamma$-ray flux according to its SFR. None of these two galaxies appear as a strong candidate for VHE detection.

\paragraph{NGC\,3621, NGC\,55, M81, NGC\,6946, M101, IC\,342 and NGC\,300 (see Fig.~\ref{fig: SED_other}):} 

For these galaxies, the extrapolated upper limit and the empirical models are consistent, with the exception of NGC\,6946, where the empirical spectrum prediction is marginally above the UL. This discrepancy  between the two models for NGC\,6946 may be attributed to its location near the Galactic plane ($b \approx 11.7$\,deg), where the \textit{Fermi}-LAT background model is less accurate, or to the relatively large uncertainty on its SFR estimate. In all cases, the spectra are below the sensitivity curves of TeV observatories. Earlier studies of SFGs \citep{Shimono2021} suggested that IC\,342 and NGC\,6946 are likely to be detected at VHE. This is at variance with our conclusions, which highlights the importance of our updated study of \textit{Fermi}-LAT data and of our comparison with the appropriate instrument response functions at higher energies.

\section{Conclusion}

\label{sec: conc}

SFGs are of significant interest due to their association with young stars and supernovae, which enable particle acceleration and CR production. Their high target density and strong magnetic fields foster CR interactions, resulting in non-thermal radiation, including $\gamma$-rays. However, current instruments have not clearly unveiled the origin of the $\gamma$-rays from SFGs. 
In this paper, we explored SFGs detectable at VHE, using the latest \textit{Fermi}-LAT data and the expected sensitivity of current and future VHE observatories. We made use of updated data to gather distances, SFRs and $\gamma$-ray fluxes or upper limits for 27 SFGs. Using the 14 SFGs detected in the 4FGL-DR4, we present a revised $L_{\gamma, \mathrm{2 GeV}}$--SFR correlation.  We employed a simple empirical emission model to predict the $\gamma$-ray flux of SFGs based on this correlation. This model assumes diffuse emission from proton-proton interactions, with a photon index of 2.2 from GeV to TeV energies, and a $\gamma$-ray luminosity proportional to the SFR. Additionally, we constrain the $\gamma$-ray emission of these galaxies using a simple extrapolation of the \textit{Fermi}-LAT data, which allowed us to constrain their TeV emission more precisely. In both cases, we accounted for the EBL absorption at the highest energies, but did not account for internal absorption effects, as the latter depend on the specific geometry of both the CR emission region and the target photon field.

We obtained promising results regarding the prospects for detection with the CTAO. Among the fourteen GeV-detected SFGs, we identified three new candidates,  NGC\,1068, NGC\,4945, and Circinus -- in addition to M82, NGC\,253 and the Magellanic Clouds -- that we recommend for observation in the VHE domain. 
Notably, two galaxies that are not currently cataloged at GeV energies, M83 and M33, may also be within the CTAO detection range, highlighting their potential as targets for further investigation. 

The case of M\,31 is more open in that, while its SFR suggests detectability at VHE, the extrapolation of its uncertain spectrum at lower energies does not. We found that galaxies such as Arp220, Arp299, NGC\,3424, and NGC\,2403 may be close to the detection limit of the CTAO. This suggests that, with an exposure time exceeding 50 hours, these galaxies have the potential to be detected in the TeV range. 

With a similar aim, \cite{Shimono2021} predicted the $\gamma$-ray emission properties of nearby galaxies based on the model of \cite{Sudoh2018} and discussed the prospect for detectability with the CTAO. With respect to this work, we confirm the relevance of NGC\,253, M82, M83 (aka NGC\,5236) and M33 but discard NGC\, 6946, and IC\,342 as primary targets. We also highlight three new candidates suggested for observation with the CTAO (NGC\,1068, NGC\,4945, and the Circinus galaxy) and list four other galaxies that are near the CTAO sensitivity limit (NGC\,2403, NGC\,3424, Arp\,220, and Arp\,299).

This work is also one of the first to conduct a straightforward population analysis of SFG candidates for study by SWGO and LHAASO. NGC\,253 and M82 have the best potential for detection by LHAASO, while NGC\,2403 may not be too far from the detection limit, provided their GeV flux can be extrapolated to the highest energies. NGC\,253, NGC\,4945 and the Circinus galaxy could potentially be detected by the upcoming SWGO experiment. 
These galaxies may nonetheless remain undetectable due to the effects of internal absorption at energies greater than 10\,TeV. A detailed model of their radiation field intensity and spatial distribution, along with VHE observations, might offer significant insights into constraining the impact of internal absorption on their spectrum. We recommend detailed predictions of VHE spectra for the identified candidates to fully assess their detection potential. Additionally, improved measurements of the GeV spectra of M31, M33, M83, Arp\,220 and Arp\,299 would provide a better understanding of their emission and a sharper view of the prospects for detection at VHE.

In a multimessenger context, investigating nearby SFGs with forthcoming observations presents a unique opportunity to constrain their neutrino flux \citep{Ambrosone2021}. Comprehensive research of individual neutrino emission of SFGs and their collective impact on the neutrino background could shed light on the origin of TeV neutrinos from NGC\,1068 \citep{IceCube2022}. Upcoming neutrino telescopes will potentially detect other point-like sources colocated with nearby SFGs, linking emissions to observed star-forming or black-hole activity.

In conclusion, our study offers a new perspective on potential VHE sources among SFGs, paving the way for future observations and advancements in understanding high-energy astrophysical phenomena in these galaxies. Soon, the CTAO will be ready to investigate the production and propagation of CRs in various sources. Preparing both theoretical models and a large sample of $\gamma$-ray-emitting SFGs in the GeV–TeV range is essential to test these predictions. Having identified the most promising candidates, further investigation of these galaxies is warranted. More precise predictions of their spectra, achieved through CR transport modeling and multiwavelength spectral fitting, would enhance our understanding and predictive capabilities for their TeV emissions.



\begin{acknowledgements}
This research has made use of the SIMBAD database, operated at CDS, Strasbourg, France, the NASA's Astrophysics Data System (ADS) and also the CTAO instrument response functions provided by the CTAO Consortium and Observatory, see https://www.ctao-observatory.org/science/cta-performance/  \citep[version prod5 v0.1;][for more details]{CTAsens}. The authors would like to thank the reviewer, whose comments substantially improved the quality of this manuscript. PK acknowledges financial support from the Severo Ochoa grant CEX2021-001131-S funded by MCIN/AEI/ 10.13039/501100011033.
JB gratefully acknowledges funding from ANR via the grant MICRO, ANR-20-CE92-0052. 
\end{acknowledgements}
%
\bibliographystyle{aa} 
\bibliography{ref} 

\begin{thebibliography}{102}
\expandafter\ifx\csname natexlab\endcsname\relax\def\natexlab#1{#1}\fi

\bibitem[{{Abdo} {et~al.}(2010{\natexlab{a}}){Abdo}, {Ackermann}, {Ajello},
  {Allafort}, {Antolini}, {Atwood}, {Axelsson}, {Baldini}, {Ballet},
  {Barbiellini}, {Bastieri}, {Baughman}, {Bechtol}, {Bellazzini}, {Belli},
  {Berenji}, {Bisello}, {Blandford}, {Bloom}, {Bonamente}, {Bonnell},
  {Borgland}, {Bouvier}, {Bregeon}, {Brez}, {Brigida}, {Bruel}, {Burnett},
  {Busetto}, {Buson}, {Caliandro}, {Cameron}, {Campana}, {Canadas}, {Caraveo},
  {Carrigan}, {Casandjian}, {Cavazzuti}, {Ceccanti}, {Cecchi}, {{\c{C}}elik},
  {Charles}, {Chekhtman}, {Cheung}, {Chiang}, {Cillis}, {Ciprini}, {Claus},
  {Cohen-Tanugi}, {Conrad}, {Corbet}, {Davis}, {DeKlotz}, {den Hartog},
  {Dermer}, {de Angelis}, {de Luca}, {de Palma}, {Digel}, {Dormody}, {Silva},
  {Drell}, {Dubois}, {Dumora}, {Fabiani}, {Farnier}, {Favuzzi}, {Fegan},
  {Ferrara}, {Focke}, {Fortin}, {Frailis}, {Fukazawa}, {Funk}, {Fusco},
  {Gargano}, {Gasparrini}, {Gehrels}, {Germani}, {Giavitto}, {Giebels},
  {Giglietto}, {Giommi}, {Giordano}, {Giroletti}, {Glanzman}, {Godfrey},
  {Grenier}, {Grondin}, {Grove}, {Guillemot}, {Guiriec}, {Gustafsson},
  {Hadasch}, {Hanabata}, {Harding}, {Hayashida}, {Hays}, {Healey}, {Hill},
  {Horan}, {Hughes}, {Iafrate}, {J{\'o}hannesson}, {Johnson}, {Johnson},
  {Johnson}, {Johnson}, {Kamae}, {Katagiri}, {Kataoka}, {Kawai}, {Kerr},
  {Kn{\"o}dlseder}, {Kocevski}, {Kuss}, {Lande}, {Landriu}, {Latronico}, {Lee},
  {Lemoine-Goumard}, {Lionetto}, {Llena Garde}, {Longo}, {Loparco}, {Lott},
  {Lovellette}, {Lubrano}, {Madejski}, {Makeev}, {Marangelli}, {Marelli},
  {Massaro}, {Mazziotta}, {McConville}, {McEnery}, {Michelson}, {Minuti},
  {Mitthumsiri}, {Mizuno}, {Moiseev}, {Mongelli}, {Monte}, {Monzani},
  {Moretti}, {Morselli}, {Moskalenko}, {Murgia}, {Nakajima}, {Nakamori},
  {Naumann-Godo}, {Nolan}, {Norris}, {Nuss}, {Ohno}, {Ohsugi}, {Omodei},
  {Orlando}, {Ormes}, {Ozaki}, {Paccagnella}, {Paneque}, {Panetta}, {Parent},
  {Pelassa}, {Pepe}, {Pesce-Rollins}, {Pinchera}, {Piron}, {Porter}, {Poupard},
  {Rain{\`o}}, {Rando}, {Ray}, {Razzano}, {Razzaque}, {Rea}, {Reimer},
  {Reimer}, {Reposeur}, {Ripken}, {Ritz}, {Rochester}, {Rodriguez}, {Romani},
  {Roth}, {Sadrozinski}, {Salvetti}, {Sanchez}, {Sander}, {Saz Parkinson},
  {Scargle}, {Schalk}, {Scolieri}, {Sgr{\`o}}, {Shaw}, {Siskind}, {Smith},
  {Smith}, {Spandre}, {Spinelli}, {Starck}, {Stephens}, {Striani}, {Strickman},
  {Strong}, {Suson}, {Tajima}, {Takahashi}, {Takahashi}, {Tanaka}, {Thayer},
  {Thayer}, {Thompson}, {Tibaldo}, {Tibolla}, {Tinebra}, {Torres}, {Tosti},
  {Tramacere}, {Uchiyama}, {Usher}, {Van Etten}, {Vasileiou}, {Vilchez},
  {Vitale}, {Waite}, {Wallace}, {Wang}, {Watters}, {Winer}, {Wood}, {Yang},
  {Ylinen}, {Ziegler}, \& {Fermi LAT Collaboration}}]{1FGL}
{Abdo}, A.~A., {Ackermann}, M., {Ajello}, M., {et~al.} 2010{\natexlab{a}},
  \apjs, 188, 405

\bibitem[{{Abdo} {et~al.}(2010{\natexlab{b}}){Abdo}, {Ackermann}, {Ajello},
  {Allafort}, {Atwood}, {Baldini}, {Ballet}, {Barbiellini}, {Bastieri},
  {Bechtol}, {Bellazzini}, {Berenji}, {Blandford}, {Bloom}, {Bonamente},
  {Borgland}, {Bouvier}, {Brandt}, {Bregeon}, {Brigida}, {Bruel}, {Buehler},
  {Burnett}, {Buson}, {Caliandro}, {Cameron}, {Cannon}, {Caraveo},
  {Casandjian}, {Cecchi}, {{\c{C}}elik}, {Charles}, {Chekhtman}, {Chiang},
  {Ciprini}, {Claus}, {Cohen-Tanugi}, {Conrad}, {Dermer}, {de Angelis}, {de
  Palma}, {Digel}, {Silva}, {Drell}, {Drlica-Wagner}, {Dubois}, {Favuzzi},
  {Fegan}, {Fortin}, {Frailis}, {Fukazawa}, {Funk}, {Fusco}, {Gargano},
  {Germani}, {Giglietto}, {Giordano}, {Giroletti}, {Glanzman}, {Godfrey},
  {Grenier}, {Grondin}, {Guiriec}, {Gustafsson}, {Hadasch}, {Harding},
  {Hayashi}, {Hayashida}, {Hays}, {Healey}, {Jean}, {J{\'o}hannesson},
  {Johnson}, {Johnson}, {Johnson}, {Kamae}, {Katagiri}, {Kataoka}, {Kerr},
  {Kn{\"o}dlseder}, {Kuss}, {Lande}, {Latronico}, {Lee}, {Lemoine-Goumard},
  {Longo}, {Loparco}, {Lott}, {Lovellette}, {Lubrano}, {Madejski}, {Makeev},
  {Martin}, {Mazziotta}, {Mehault}, {Michelson}, {Mitthumsiri}, {Mizuno},
  {Moiseev}, {Monte}, {Monzani}, {Morselli}, {Moskalenko}, {Murgia},
  {Naumann-Godo}, {Nolan}, {Norris}, {Nuss}, {Ohsugi}, {Okumura}, {Omodei},
  {Orlando}, {Ormes}, {Ozaki}, {Paneque}, {Panetta}, {Parent}, {Pepe},
  {Persic}, {Pesce-Rollins}, {Piron}, {Porter}, {Rain{\`o}}, {Rando},
  {Razzano}, {Reimer}, {Reimer}, {Ritz}, {Romani}, {Sadrozinski}, {Saz
  Parkinson}, {Sgr{\`o}}, {Siskind}, {Smith}, {Smith}, {Spandre}, {Spinelli},
  {Strickman}, {Strigari}, {Strong}, {Suson}, {Takahashi}, {Takahashi},
  {Tanaka}, {Thayer}, {Thompson}, {Tibaldo}, {Torres}, {Tosti}, {Tramacere},
  {Uchiyama}, {Usher}, {Vandenbroucke}, {Vianello}, {Vilchez}, {Vitale},
  {Waite}, {Wang}, {Winer}, {Wood}, {Yang}, \& {Ziegler}}]{Abdo2010}
{Abdo}, A.~A., {Ackermann}, M., {Ajello}, M., {et~al.} 2010{\natexlab{b}},
  \aap, 523, L2

\bibitem[{{Abramowski} {et~al.}(2012){Abramowski}, {Acero}, {Aharonian},
  {Akhperjanian}, {Anton}, {Balzer}, {Barnacka}, {Becherini}, {Becker},
  {Bernl{\"o}hr}, {Birsin}, {Biteau}, {Bochow}, {Boisson}, {Bolmont}, {Bordas},
  {Brucker}, {Brun}, {Brun}, {Bulik}, {B{\"u}sching}, {Carrigan}, {Casanova},
  {Cerruti}, {Chadwick}, {Charbonnier}, {Chaves}, {Cheesebrough}, {Cologna},
  {Conrad}, {Couturier}, {Dalton}, {Daniel}, {Davids}, {Degrange}, {Deil},
  {Dickinson}, {Djannati-Ata{\"\i}}, {Domainko}, {Drury}, {Dubus}, {Dutson},
  {Dyks}, {Dyrda}, {Egberts}, {Eger}, {Espigat}, {Fallon}, {Fegan},
  {Feinstein}, {Fernandes}, {Fiasson}, {Fontaine}, {F{\"o}rster},
  {F{\"u}{\ss}ling}, {Gajdus}, {Gallant}, {Garrigoux}, {Gast}, {G{\'e}rard},
  {Giebels}, {Glicenstein}, {Gl{\"u}ck}, {G{\"o}ring}, {Grondin},
  {H{\"a}ffner}, {Hague}, {Hahn}, {Hampf}, {Harris}, {Hauser}, {Heinz},
  {Heinzelmann}, {Henri}, {Hermann}, {Hillert}, {Hinton}, {Hofmann},
  {Hofverberg}, {Holler}, {Horns}, {Jacholkowska}, {Jahn}, {Jamrozy}, {Jung},
  {Kastendieck}, {Katarzy{\'n}ski}, {Katz}, {Kaufmann}, {Kh{\'e}lifi},
  {Klochkov}, {Klu{\'z}niak}, {Kneiske}, {Komin}, {Kosack}, {Kossakowski},
  {Krayzel}, {Laffon}, {Lamanna}, {Lenain}, {Lennarz}, {Lohse}, {Lopatin},
  {Lu}, {Marandon}, {Marcowith}, {Masbou}, {Maurin}, {Maxted}, {Mayer},
  {McComb}, {Medina}, {M{\'e}hault}, {Moderski}, {Mohamed}, {Moulin},
  {Naumann}, {Naumann-Godo}, {de Naurois}, {Nedbal}, {Nekrassov}, {Nguyen},
  {Nicholas}, {Niemiec}, {Nolan}, {Ohm}, {de O{\~n}a Wilhelmi}, {Opitz},
  {Ostrowski}, {Oya}, {Panter}, {Paz Arribas}, {Pekeur}, {Pelletier}, {Perez},
  {Petrucci}, {Peyaud}, {Pita}, {P{\"u}hlhofer}, {Punch}, {Quirrenbach},
  {Raue}, {Reimer}, {Reimer}, {Renaud}, {de los Reyes}, {Rieger}, {Ripken},
  {Rob}, {Rosier-Lees}, {Rowell}, {Rudak}, {Rulten}, {Sahakian}, {Sanchez},
  {Santangelo}, {Schlickeiser}, {Schulz}, {Schwanke}, {Schwarzburg},
  {Schwemmer}, {Sheidaei}, {Skilton}, {Sol}, {Spengler}, {Stawarz},
  {Steenkamp}, {Stegmann}, {Stinzing}, {Stycz}, {Sushch}, {Szostek},
  {Tavernet}, {Terrier}, {Tluczykont}, {Valerius}, {van Eldik}, {Vasileiadis},
  {Venter}, {Viana}, {Vincent}, {V{\"o}lk}, {Volpe}, {Vorobiov}, {Vorster},
  {Wagner}, {Ward}, {White}, {Wierzcholska}, {Zacharias}, {Zajczyk},
  {Zdziarski}, {Zech}, {Zechlin}, \& {H.~E.~S.~S.
  Collaboration}}]{Abramowski2012}
{Abramowski}, A., {Acero}, F., {Aharonian}, F., {et~al.} 2012, \apj, 757, 158

\bibitem[{{Acciari} {et~al.}(2019){Acciari}, {Ansoldi}, {Antonelli}, {Arbet
  Engels}, {Baack}, {Babi{\'c}}, {Banerjee}, {Barres de Almeida}, {Barrio},
  {Becerra Gonz{\'a}lez}, {Bednarek}, {Bellizzi}, {Bernardini}, {Berti},
  {Besenrieder}, {Bhattacharyya}, {Bigongiari}, {Biland}, {Blanch}, {Bonnoli},
  {Bo{\v{s}}njak}, {Busetto}, {Carosi}, {Ceribella}, {Chai}, {Chilingaryan},
  {Cikota}, {Colak}, {Colin}, {Colombo}, {Contreras}, {Cortina}, {Covino},
  {D'Elia}, {Da Vela}, {Dazzi}, {De Angelis}, {De Lotto}, {Delfino}, {Delgado},
  {Depaoli}, {Di Pierro}, {Di Venere}, {Do Souto Espi{\~n}eira}, {Dominis
  Prester}, {Donini}, {Dorner}, {Doro}, {Elsaesser}, {Fallah Ramazani},
  {Fattorini}, {Ferrara}, {Fidalgo}, {Foffano}, {Fonseca}, {Font}, {Fruck},
  {Fukami}, {Garc{\'\i}a L{\'o}pez}, {Garczarczyk}, {Gasparyan}, {Gaug},
  {Giglietto}, {Giordano}, {Godinovi{\'c}}, {Green}, {Guberman}, {Hadasch},
  {Hahn}, {Herrera}, {Hoang}, {Hrupec}, {H{\"u}tten}, {Inada}, {Inoue},
  {Ishio}, {Iwamura}, {Jouvin}, {Kerszberg}, {Kubo}, {Kushida}, {Lamastra},
  {Lelas}, {Leone}, {Lindfors}, {Lombardi}, {Longo}, {L{\'o}pez},
  {L{\'o}pez-Coto}, {L{\'o}pez-Oramas}, {Loporchio}, {Machado de Oliveira
  Fraga}, {Maggio}, {Majumdar}, {Makariev}, {Mallamaci}, {Maneva}, {Manganaro},
  {Mannheim}, {Maraschi}, {Mariotti}, {Mart{\'\i}nez}, {Mazin},
  {Mi{\'c}anovi{\'c}}, {Miceli}, {Minev}, {Miranda}, {Mirzoyan}, {Molina},
  {Moralejo}, {Morcuende}, {Moreno}, {Moretti}, {Munar-Adrover}, {Neustroev},
  {Nigro}, {Nilsson}, {Ninci}, {Nishijima}, {Noda}, {Nogu{\'e}s}, {Nozaki},
  {Paiano}, {Palacio}, {Palatiello}, {Paneque}, {Paoletti}, {Paredes},
  {Pe{\~n}il}, {Peresano}, {Persic}, {Prada Moroni}, {Prandini}, {Puljak},
  {Rhode}, {Rib{\'o}}, {Rico}, {Righi}, {Rugliancich}, {Saha}, {Sahakyan},
  {Saito}, {Sakurai}, {Satalecka}, {Schmidt}, {Schweizer}, {Sitarek},
  {{\v{S}}nidari{\'c}}, {Sobczynska}, {Somero}, {Stamerra}, {Strom}, {Strzys},
  {Suda}, {Suri{\'c}}, {Takahashi}, {Tavecchio}, {Temnikov}, {Terzi{\'c}},
  {Teshima}, {Torres-Alb{\`a}}, {Tosti}, {Vagelli}, {van Scherpenberg},
  {Vanzo}, {Vazquez Acosta}, {Vigorito}, {Vitale}, {Vovk}, {Will}, {Zari{\'c}},
  {MAGIC Collaboration}, {Fiore}, {Feruglio}, \& {Rephaeli}}]{MAGIC2019}
{Acciari}, V.~A., {Ansoldi}, S., {Antonelli}, L.~A., {et~al.} 2019, \apj, 883,
  135

\bibitem[{{Acharyya} {et~al.}(2023){Acharyya}, {Adam}, {Aguasca-Cabot},
  {Agudo}, {Aguirre-Santaella}, {Alfaro}, {Aloisio}, {Alves Batista}, {Amato},
  {Ang{\"u}ner}, {Aramo}, {Arcaro}, {Asano}, {Aschersleben}, {Ashkar},
  {Backes}, {Baktash}, {Balazs}, {Balbo}, {Ballet}, {Bamba}, {Baquero Larriva},
  {Barbosa Martins}, {Barres de Almeida}, {Barrio}, {Bastieri}, {Batista},
  {Batkovic}, {Baxter}, {Becerra Gonz{\'a}lez}, {Becker Tjus}, {Benbow},
  {Bernardini}, {Bernardos Mart{\'\i}n}, {Bernete Medrano}, {Berti},
  {Bertucci}, {Beshley}, {Bhattacharjee}, {Bhattacharyya}, {Bigongiari},
  {Biland}, {Bissaldi}, {Bocchino}, {Bordas}, {Borkowski}, {Bottacini},
  {B{\"o}ttcher}, {Bradascio}, {Brown}, {Bulgarelli}, {Burmistrov}, {Caroff},
  {Carosi}, {Carqu{\'\i}n}, {Casanova}, {Cascone}, {Cassol}, {Cerruti},
  {Chadwick}, {Chaty}, {Chen}, {Chiavassa}, {Chytka}, {Conforti}, {Cortina},
  {Costa}, {Costantini}, {Cotter}, {Crestan}, {Cristofari}, {D'Ammando},
  {Dalchenko}, {Dazzi}, {De Angelis}, {De Caprio}, {de Gouveia Dal Pino}, {De
  Martino}, {de Naurois}, {de Souza}, {del Valle}, {Delgado Giler}, {Delgado},
  {della Volpe}, {Depaoli}, {Di Girolamo}, {Di Piano}, {Di Pierro}, {Di Tria},
  {Di Venere}, {Diebold}, {Doro}, {Dumora}, {Dwarkadas}, {Eckner}, {Egberts},
  {Emery}, {Escudero}, {Falceta-Goncalves}, {Fedorova}, {Fegan}, {Feng},
  {Ferenc}, {Ferrand}, {Fiandrini}, {Filipovic}, {Fioretti}, {Foffano},
  {Fontaine}, {Fukui}, {Gaggero}, {Galanti}, {Galaz}, {Gallozzi}, {Gammaldi},
  {Garczarczyk}, {Gasbarra}, {Gasparrini}, {Ghalumyan}, {Giarrusso},
  {Giavitto}, {Giglietto}, {Giordano}, {Giuliani}, {Glicenstein}, {Goldoni},
  {Goulart Coelho}, {Granot}, {Green}, {Green}, {Grondin}, {Gueta}, {Hadasch},
  {Hamal}, {Hassan}, {Hayashi}, {Heller}, {Hern{\'a}ndez Cadena}, {Hiroshima},
  {Hnatyk}, {Hnatyk}, {Hofmann}, {Holder}, {Holler}, {Horan}, {Horvath},
  {Hrabovsky}, {H{\"u}tten}, {Iarlori}, {Inada}, {Incardona}, {Inoue}, {Iocco},
  {Jamrozy}, {Jin}, {Jung-Richardt}, {Jury{\v{s}}ek}, {Kantzas}, {Karas},
  {Katagiri}, {Kerszberg}, {Kn{\"o}dlseder}, {Komin}, {Kornecki}, {Kosack},
  {Kowal}, {Kubo}, {Lamastra}, {Lapington}, {Lemoine-Goumard}, {Lenain},
  {Leone}, {Leto}, {Leuschner}, {Lindfors}, {Lohse}, {Lombardi}, {Longo},
  {L{\'o}pez-Coto}, {L{\'o}pez-Oramas}, {Loporchio}, {Luque-Escamilla},
  {Macias}, {Majumdar}, {Mandat}, {Mangano}, {Manic{\`o}}, {Mariotti},
  {Marquez}, {Marsella}, {Mart{\'\i}}, {Martin}, {Mart{\'\i}nez}, {Mazin},
  {Menchiari}, {Meyer}, {Miceli}, {Miceli}, {Micha{\l}owski}, {Mitchell},
  {Moderski}, {Mohrmann}, {Molero}, {Molina}, {Montaruli}, {Moralejo},
  {Morcuende}, {Morselli}, {Moulin}, {Moya}, {Mukherjee}, {Munari},
  {Muraczewski}, {Nagataki}, {Nakamori}, {Nayak}, {Niemiec}, {Nievas},
  {Niko{\l}ajuk}, {Nishijima}, {Noda}, {Nosek}, {Novosyadlyj}, {Nozaki},
  {Ohishi}, {Ohm}, {Okumura}, {Olmi}, {Ong}, {Orienti}, {Orito}, {Orlandini},
  {Orlando}, {Orlando}, {Ostrowski}, {Oya}, {Pagliaro}, {Palatka}, {Pantaleo},
  {Paoletti}, {Paredes}, {Parmiggiani}, {Patricelli}, {Pech}, {Pecimotika},
  {Persic}, {Petruk}, {Pierre}, {Pietropaolo}, {Pirola}, {Pohl}, {Prandini},
  {Priyadarshi}, {P{\"u}hlhofer}, {Pumo}, {Punch}, {Queiroz}, {Quirrenbach},
  {Rain{\`o}}, {Rando}, {Razzaque}, {Reimer}, {Reimer}, {Reposeur}, {Rib{\'o}},
  {Richtler}, {Rico}, {Rieger}, {Rigoselli}, {Rizi}, {Roache}, {Rodriguez
  Fernandez}, {Romano}, {Romeo}, {Rosado}, {Rosales de Leon}, {Rudak},
  {Rulten}, {Sadeh}, {Saito}, {S{\'a}nchez-Conde}, {Sano}, {Santangelo},
  {Santos-Lima}, {Sarkar}, {Saturni}, {Scherer}, {Schovanek}, {Schussler},
  {Schwanke}, {Sergijenko}, {Servillat}, {Siejkowski}, {Siqueira}, {Spencer},
  {Stamerra}, {Stani{\v{c}}}, {Steppa}, {Stolarczyk}, {Suda}, {Tavernier},
  {Teshima}, {Tibaldo}, {Torres}, {Tothill}, {Vacula}, {Vallage}, {Vallania},
  {van Eldik}, {V{\'a}zquez Acosta}, {Vecchi}, {Ventura}, {Vercellone},
  {Viana}, {Vigorito}, {Vink}, {Vitale}, {Vodeb}, {Vorobiov}, {Vuillaume},
  {Wagner}, {Walter}, {White}, {Wierzcholska}, {Will}, {Yamazaki}, {Yang},
  {Yoshikoshi}, {Zacharias}, {Zaharijas}, {Zavrtanik}, {Zavrtanik},
  {Zdziarski}, {Zhdanov}, {Zi{\k{e}}tara}, \& {{\v{Z}}ivec}}]{LMCCTA2023}
{Acharyya}, A., {Adam}, R., {Aguasca-Cabot}, A., {et~al.} 2023, \mnras, 523,
  5353

\bibitem[{{Ackermann} {et~al.}(2017){Ackermann}, {Ajello}, {Albert}, {Baldini},
  {Ballet}, {Barbiellini}, {Bastieri}, {Bellazzini}, {Bissaldi}, {Bloom},
  {Bonino}, {Bottacini}, {Brandt}, {Bregeon}, {Bruel}, {Buehler}, {Cameron},
  {Caputo}, {Caragiulo}, {Caraveo}, {Cavazzuti}, {Cecchi}, {Charles},
  {Chekhtman}, {Chiaro}, {Ciprini}, {Costanza}, {Cutini}, {D'Ammando}, {de
  Palma}, {Desiante}, {Digel}, {Di Lalla}, {Di Mauro}, {Di Venere}, {Favuzzi},
  {Funk}, {Fusco}, {Gargano}, {Giglietto}, {Giordano}, {Giroletti}, {Glanzman},
  {Green}, {Grenier}, {Guillemot}, {Guiriec}, {Hayashi}, {Hou},
  {J{\'o}hannesson}, {Kamae}, {Kn{\"o}dlseder}, {Kong}, {Kuss}, {La Mura},
  {Larsson}, {Latronico}, {Li}, {Longo}, {Loparco}, {Lubrano}, {Maldera},
  {Malyshev}, {Manfreda}, {Martin}, {Mazziotta}, {Michelson}, {Mirabal},
  {Mitthumsiri}, {Mizuno}, {Monzani}, {Morselli}, {Moskalenko}, {Negro},
  {Nuss}, {Ohsugi}, {Omodei}, {Orlando}, {Ormes}, {Paneque}, {Persic},
  {Pesce-Rollins}, {Piron}, {Porter}, {Principe}, {Rain{\`o}}, {Rando},
  {Razzano}, {Reimer}, {S{\'a}nchez-Conde}, {Sgr{\`o}}, {Simone}, {Siskind},
  {Spada}, {Spandre}, {Spinelli}, {Tanaka}, {Tibaldo}, {Torres}, {Troja},
  {Uchiyama}, {Wang}, {Wood}, {Wood}, {Zaharijas}, \& {Zhou}}]{Ackermann2017}
{Ackermann}, M., {Ajello}, M., {Albert}, A., {et~al.} 2017, \apj, 836, 208

\bibitem[{{Ackermann} {et~al.}(2012){Ackermann}, {Ajello}, {Allafort},
  {Baldini}, {Ballet}, {Bastieri}, {Bechtol}, {Bellazzini}, {Berenji}, {Bloom},
  {Bonamente}, {Borgland}, {Bouvier}, {Bregeon}, {Brigida}, {Bruel}, {Buehler},
  {Buson}, {Caliandro}, {Cameron}, {Caraveo}, {Casandjian}, {Cecchi},
  {Charles}, {Chekhtman}, {Cheung}, {Chiang}, {Cillis}, {Ciprini}, {Claus},
  {Cohen-Tanugi}, {Conrad}, {Cutini}, {de Palma}, {Dermer}, {Digel}, {Silva},
  {Drell}, {Drlica-Wagner}, {Favuzzi}, {Fegan}, {Fortin}, {Fukazawa}, {Funk},
  {Fusco}, {Gargano}, {Gasparrini}, {Germani}, {Giglietto}, {Giordano},
  {Glanzman}, {Godfrey}, {Grenier}, {Guiriec}, {Gustafsson}, {Hadasch},
  {Hayashida}, {Hays}, {Hughes}, {J{\'o}hannesson}, {Johnson}, {Kamae},
  {Katagiri}, {Kataoka}, {Kn{\"o}dlseder}, {Kuss}, {Lande}, {Longo}, {Loparco},
  {Lott}, {Lovellette}, {Lubrano}, {Madejski}, {Martin}, {Mazziotta},
  {McEnery}, {Michelson}, {Mizuno}, {Monte}, {Monzani}, {Morselli},
  {Moskalenko}, {Murgia}, {Nishino}, {Norris}, {Nuss}, {Ohno}, {Ohsugi},
  {Okumura}, {Omodei}, {Orlando}, {Ozaki}, {Parent}, {Persic}, {Pesce-Rollins},
  {Petrosian}, {Pierbattista}, {Piron}, {Pivato}, {Porter}, {Rain{\`o}},
  {Rando}, {Razzano}, {Reimer}, {Reimer}, {Ritz}, {Roth}, {Sbarra}, {Sgr{\`o}},
  {Siskind}, {Spandre}, {Spinelli}, {Stawarz}, {Strong}, {Takahashi}, {Tanaka},
  {Thayer}, {Tibaldo}, {Tinivella}, {Torres}, {Tosti}, {Troja}, {Uchiyama},
  {Vandenbroucke}, {Vianello}, {Vitale}, {Waite}, {Wood}, \&
  {Yang}}]{Ackermann2012}
{Ackermann}, M., {Ajello}, M., {Allafort}, A., {et~al.} 2012, \apj, 755, 164

\bibitem[{{Ajello} {et~al.}(2020){Ajello}, {Di Mauro}, {Paliya}, \&
  {Garrappa}}]{Ajello2020}
{Ajello}, M., {Di Mauro}, M., {Paliya}, V.~S., \& {Garrappa}, S. 2020, \apj,
  894, 88

\bibitem[{{Albert} {et~al.}(2019){Albert}, {Alfaro}, {Ashkar}, {Alvarez},
  {{\'A}lvarez}, {Arteaga-Vel{\'a}zquez}, {Ayala Solares}, {Arceo}, {Bellido},
  {BenZvi}, {Bretz}, {Brisbois}, {Brown}, {Brun}, {Caballero-Mora}, {Carosi},
  {Carrami{\~n}ana}, {Casanova}, {Chadwick}, {Cotter}, {Couti{\~n}o De
  Le{\'o}n}, {Cristofari}, {Dasso}, {de la Fuente}, {Dingus}, {Desiati},
  {Salles}, {de Souza}, {Dorner}, {D{\'\i}az-V{\'e}lez},
  {Garc{\'\i}a-Gonz{\'a}lez}, {DuVernois}, {Di Sciascio}, {Engel},
  {Fleischhack}, {Fraija}, {Funk}, {Glicenstein}, {Gonzalez}, {Gonz{\'a}lez},
  {Goodman}, {Harding}, {Haungs}, {Hinton}, {Hona}, {Hoyos}, {Huentemeyer},
  {Iriarte}, {Jardin-Blicq}, {Joshi}, {Kaufmann}, {Kawata}, {Kunwar},
  {Lefaucheur}, {Lenain}, {Link}, {L{\'o}pez-Coto}, {Marandon}, {Mariotti},
  {Mart{\'\i}nez-Castro}, {Mart{\'\i}nez-Huerta}, {Mostaf{\'a}}, {Nayerhoda},
  {Nellen}, {de O{\~n}a Wilhelmi}, {Parsons}, {Patricelli}, {Pichel}, {Piel},
  {Prandini}, {Pueschel}, {Procureur}, {Reisenegger}, {Rivi{\`e}re},
  {Rodriguez}, {Rovero}, {Rowell}, {Ruiz-Velasco}, {Sandoval}, {Santander},
  {Sako}, {Sako}, {Satalecka}, {Schoorlemmer}, {Sch{\"u}ssler},
  {Seglar-Arroyo}, {Smith}, {Spencer}, {Surajbali}, {Tabachnick}, {Taylor},
  {Tibolla}, {Torres}, {Vallage}, {Viana}, {Watson}, {Weisgarber}, {Werner},
  {White}, {Wischnewski}, {Yang}, {Zepeda}, \& {Zhou}}]{Albert2019}
{Albert}, A., {Alfaro}, R., {Ashkar}, H., {et~al.} 2019, arXiv e-prints,
  arXiv:1902.08429

\bibitem[{{Ambrosone} {et~al.}(2021){Ambrosone}, {Chianese}, {Fiorillo},
  {Marinelli}, \& {Miele}}]{Ambrosone2021}
{Ambrosone}, A., {Chianese}, M., {Fiorillo}, D. F.~G., {Marinelli}, A., \&
  {Miele}, G. 2021, \apjl, 919, L32

\bibitem[{{Ambrosone} {et~al.}(2024){Ambrosone}, {Chianese}, \&
  {Marinelli}}]{Ambrosone2024}
{Ambrosone}, A., {Chianese}, M., \& {Marinelli}, A. 2024, \jcap, 2024, 040

\bibitem[{{Antonucci} \& {Miller}(1985)}]{Antonucci1985}
{Antonucci}, R.~R.~J. \& {Miller}, J.~S. 1985, \apj, 297, 621

\bibitem[{{Axford} {et~al.}(1977){Axford}, {Leer}, \& {Skadron}}]{Axford1977}
{Axford}, W.~I., {Leer}, E., \& {Skadron}, G. 1977, in International Cosmic Ray
  Conference, Vol.~11, International Cosmic Ray Conference, 132

\bibitem[{{Ballet} {et~al.}(2023){Ballet}, {Bruel}, {Burnett}, {Lott}, \& {The
  Fermi-LAT collaboration}}]{4fgl}
{Ballet}, J., {Bruel}, P., {Burnett}, T.~H., {Lott}, B., \& {The Fermi-LAT
  collaboration}. 2023, arXiv e-prints, arXiv:2307.12546

\bibitem[{{Bell}(1978)}]{bell1978}
{Bell}, A.~R. 1978, \mnras, 182, 147

\bibitem[{{Bell}(2003)}]{Bell2003}
{Bell}, E.~F. 2003, \apj, 586, 794

\bibitem[{{Biteau}(2021)}]{Biteau2021}
{Biteau}, J. 2021, \apjs, 256, 15

\bibitem[{{Biteau} \& {Meyer}(2022)}]{2022Galax..10...39B}
{Biteau}, J. \& {Meyer}, M. 2022, Galaxies, 10, 39

\bibitem[{{Blom} {et~al.}(1999){Blom}, {Paglione}, \&
  {Carrami{\~n}ana}}]{Blom1999}
{Blom}, J.~J., {Paglione}, T. A.~D., \& {Carrami{\~n}ana}, A. 1999, \apj, 516,
  744

\bibitem[{{Bouquin} {et~al.}(2018){Bouquin}, {Gil de Paz}, {Mu{\~n}oz-Mateos},
  {Boissier}, {Sheth}, {Zaritsky}, {Peletier}, {Knapen}, \&
  {Gallego}}]{Bouquin2018}
{Bouquin}, A. Y.~K., {Gil de Paz}, A., {Mu{\~n}oz-Mateos}, J.~C., {et~al.}
  2018, \apjs, 234, 18

\bibitem[{{Bruzewski} {et~al.}(2023){Bruzewski}, {Schinzel}, \&
  {Taylor}}]{Bruzewski2023}
{Bruzewski}, S., {Schinzel}, F.~K., \& {Taylor}, G.~B. 2023, \apj, 943, 51

\bibitem[{{Bykov} \& {Fleishman}(1992)}]{Bykov1992}
{Bykov}, A.~M. \& {Fleishman}, G.~D. 1992, \mnras, 255, 269

\bibitem[{{Cao} {et~al.}(2019){Cao}, {della Volpe}, {Liu}, {Editors}, {:},
  {Bi}, {Chen}, {D'Ettorre Piazzoli}, {Feng}, {Jia}, {Li}, {Ma}, {Wang},
  {Zhang}, {Referees}, {:}, {Qie}, {Hu}, {Referees}, {:}, {S{\'a}iz}, {Yang},
  {Contributors}, {:}, {Addazi}, {Belotsky}, {Beylin}, {Bi}, {Che}, {Chen},
  {Cheng}, {Chiavassa}, {Cirelli}, {Di Sciascio}, {Esmaili}, {Fang},
  {Fornengo}, {Gou}, {Guo}, {Gan}, {Gong}, {Gu}, {He}, {He}, {Hou}, {Huang},
  {Huang}, {Kachekriess}, {Khlopov}, {Korchagin}, {Korochkin}, {Kuksa},
  {Ksenofontov}, {Liu}, {Liu}, {Liu}, {Marciano}, {Martineau-Huynh},
  {Martraire}, {Ma}, {Neronov}, {Panci}, {Pasechnick}, {Ruffolo}, {Sakharov},
  {Sala}, {Semikoz}, {Shchegolev}, {Serpico}, {Sheng}, {Stenkin}, {Tam},
  {Vernetto}, {Vallania}, {Volchanskiy}, {Wang}, {Wang}, {Wang}, {Wu}, {Wu},
  {Wu}, {Xiao}, {Yang}, {Yan}, {Yao}, {Yin}, {Yuan}, {Zhang}, {Zeng}, {Zhang},
  {Zhang}, {Zhou}, {Zhu}, \& {Zuo}}]{LHAASO2019}
{Cao}, Z., {della Volpe}, D., {Liu}, S., {et~al.} 2019, arXiv e-prints,
  arXiv:1905.02773

\bibitem[{{Chabrier}(2003)}]{Chabrier2003}
{Chabrier}, G. 2003, \pasp, 115, 763

\bibitem[{{Chen} {et~al.}(2024){Chen}, {Liu}, {Wang}, \& {Chang}}]{Chen2024}
{Chen}, X.-B., {Liu}, R.-Y., {Wang}, X.-Y., \& {Chang}, X.-C. 2024, \mnras,
  527, 7915

\bibitem[{{Cherenkov Telescope Array Consortium} {et~al.}(2019){Cherenkov
  Telescope Array Consortium}, {Acharya}, {Agudo}, {Al Samarai}, {Alfaro},
  {Alfaro}, {Alispach}, {Alves Batista}, {Amans}, {Amato}, {Ambrosi},
  {Antolini}, {Antonelli}, {Aramo}, {Araya}, {Armstrong}, {Arqueros},
  {Arrabito}, {Asano}, {Ashley}, {Backes}, {Balazs}, {Balbo}, {Ballester},
  {Ballet}, {Bamba}, {Barkov}, {Barres de Almeida}, {Barrio}, {Bastieri},
  {Becherini}, {Belfiore}, {Benbow}, {Berge}, {Bernardini}, {Bernardini},
  {Bernardos}, {Bernl{\"o}hr}, {Bertucci}, {Biasuzzi}, {Bigongiari}, {Biland},
  {Bissaldi}, {Biteau}, {Blanch}, {Blazek}, {Boisson}, {Bolmont}, {Bonanno},
  {Bonardi}, {Bonavolont{\`a}}, {Bonnoli}, {Bosnjak}, {B{\"o}ttcher},
  {Braiding}, {Bregeon}, {Brill}, {Brown}, {Brun}, {Brunetti}, {Buanes},
  {Buckley}, {Bugaev}, {B{\"u}hler}, {Bulgarelli}, {Bulik}, {Burton},
  {Burtovoi}, {Busetto}, {Canestrari}, {Capalbi}, {Capitanio}, {Caproni},
  {Caraveo}, {C{\'a}rdenas}, {Carlile}, {Carosi}, {Carqu{\'\i}n}, {Carr},
  {Casanova}, {Cascone}, {Catalani}, {Catalano}, {Cauz}, {Cerruti}, {Chadwick},
  {Chaty}, {Chaves}, {Chen}, {Chen}, {Chernyakova}, {Chikawa}, {Christov},
  {Chudoba}, {Cie{\'s}lar}, {Coco}, {Colafrancesco}, {Colin}, {Conforti},
  {Connaughton}, {Conrad}, {Contreras}, {Cortina}, {Costa}, {Costantini},
  {Cotter}, {Covino}, {Crocker}, {Cuadra}, {Cuevas}, {Cumani}, {D'A{\`\i}},
  {D'Ammando}, {D'Avanzo}, {D'Urso}, {Daniel}, {Davids}, {Dawson}, {Dazzi}, {De
  Angelis}, {de C{\'a}ssia dos Anjos}, {De Cesare}, {De Franco}, {de Gouveia
  Dal Pino}, {de la Calle}, {de los Reyes Lopez}, {De Lotto}, {De Luca}, {De
  Lucia}, {de Naurois}, {de O{\~n}a Wilhelmi}, {De Palma}, {De Persio}, {de
  Souza}, {Deil}, {Del Santo}, {Delgado}, {della Volpe}, {Di Girolamo}, {Di
  Pierro}, {Di Venere}, {D{\'\i}az}, {Dib}, {Diebold}, {Djannati-Ata{\"\i}},
  {Dom{\'\i}nguez}, {Dominis Prester}, {Dorner}, {Doro}, {Drass}, {Dravins},
  {Dubus}, {Dwarkadas}, {Ebr}, {Eckner}, {Egberts}, {Einecke}, {Ekoume},
  {Els{\"a}sser}, {Ernenwein}, {Espinoza}, {Evoli}, {Fairbairn},
  {Falceta-Goncalves}, {Falcone}, {Farnier}, {Fasola}, {Fedorova}, {Fegan},
  {Fernandez-Alonso}, {Fern{\'a}ndez-Barral}, {Ferrand}, {Fesquet},
  {Filipovic}, {Fioretti}, {Fontaine}, {Fornasa}, {Fortson}, {Freixas
  Coromina}, {Fruck}, {Fujita}, {Fukazawa}, {Funk}, {F{\"u}{\ss}ling},
  {Gabici}, {Gadola}, {Gallant}, {Garcia}, {Garcia L{\'o}pez}, {Garczarczyk},
  {Gaskins}, {Gasparetto}, {Gaug}, {Gerard}, {Giavitto}, {Giglietto}, {Giommi},
  {Giordano}, {Giro}, {Giroletti}, {Giuliani}, {Glicenstein}, {Gnatyk},
  {Godinovic}, {Goldoni}, {G{\'o}mez-Vargas}, {Gonz{\'a}lez}, {Gonz{\'a}lez},
  {G{\"o}tz}, {Graham}, {Grandi}, {Granot}, {Green}, {Greenshaw}, {Griffiths},
  {Gunji}, {Hadasch}, {Hara}, {Hardcastle}, {Hassan}, {Hayashi}, {Hayashida},
  {Heller}, {Helo}, {Hermann}, {Hinton}, {Hnatyk}, {Hofmann}, {Holder},
  {Horan}, {H{\"o}randel}, {Horns}, {Horvath}, {Hovatta}, {Hrabovsky},
  {Hrupec}, {Humensky}, {H{\"u}tten}, {Iarlori}, {Inada}, {Inome}, {Inoue},
  {Inoue}, {Inoue}, {Iocco}, {Ioka}, {Iori}, {Ishio}, {Iwamura}, {Jamrozy},
  {Janecek}, {Jankowsky}, {Jean}, {Jung-Richardt}, {Jurysek}, {Kaaret},
  {Karkar}, {Katagiri}, {Katz}, {Kawanaka}, {Kazanas}, {Kh{\'e}lifi}, {Kieda},
  {Kimeswenger}, {Kimura}, {Kisaka}, {Knapp}, {Kn{\"o}dlseder}, {Koch},
  {Kohri}, {Komin}, {Kosack}, {Kraus}, {Krause}, {Krau{\ss}}, {Kubo}, {Kukec
  Mezek}, {Kuroda}, {Kushida}, {La Palombara}, {Lamanna}, {Lang}, {Lapington},
  {Le Blanc}, {Leach}, {Lees}, {Lefaucheur}, {Leigui de Oliveira}, {Lenain},
  {Lico}, {Limon}, {Lindfors}, {Lohse}, {Lombardi}, {Longo}, {L{\'o}pez},
  {L{\'o}pez-Coto}, {Lu}, {Lucarelli}, {Luque-Escamilla}, {Lyard}, {Maccarone},
  {Maier}, {Majumdar}, {Malaguti}, {Mandat}, {Maneva}, {Manganaro}, {Mangano},
  {Marcowith}, {Mar{\'\i}n}, {Markoff}, {Mart{\'\i}}, {Martin},
  {Mart{\'\i}nez}, {Mart{\'\i}nez}, {Masetti}, {Masuda}, {Maurin}, {Maxted},
  {Mazin}, {Medina}, {Melandri}, {Mereghetti}, {Meyer}, {Minaya}, {Mirabal},
  {Mirzoyan}, {Mitchell}, {Mizuno}, {Moderski}, {Mohammed}, {Mohrmann},
  {Montaruli}, {Moralejo}, {Morcuende-Parrilla}, {Mori}, {Morlino}, {Morris},
  {Morselli}, {Moulin}, {Mukherjee}, {Mundell}, {Murach}, {Muraishi}, {Murase},
  {Nagai}, {Nagataki}, {Nagayoshi}, {Naito}, {Nakamori}, {Nakamura}, {Niemiec},
  {Nieto}, {Niko{\l}ajuk}, {Nishijima}, {Noda}, {Nosek}, {Novosyadlyj},
  {Nozaki}, {O'Brien}, {Oakes}, {Ohira}, {Ohishi}, {Ohm}, {Okazaki}, {Okumura},
  {Ong}, {Orienti}, {Orito}, {Osborne}, {Ostrowski}, {Otte}, {Oya}, {Padovani},
  {Paizis}, {Palatiello}, {Palatka}, {Paoletti}, {Paredes}, {Pareschi},
  {Parsons}, {Pe'er}, {Pech}, {Pedaletti}, {Perri}, {Persic}, {Petrashyk},
  {Petrucci}, {Petruk}, {Peyaud}, {Pfeifer}, {Piano}, {Pisarski}, {Pita},
  {Pohl}, {Polo}, {Pozo}, {Prandini}, {Prast}, {Principe}, {Prokhorov},
  {Prokoph}, {Prouza}, {P{\"u}hlhofer}, {Punch}, {P{\"u}rckhauer}, {Queiroz},
  {Quirrenbach}, {Rain{\`o}}, {Razzaque}, {Reimer}, {Reimer}, {Reisenegger},
  {Renaud}, {Rezaeian}, {Rhode}, {Ribeiro}, {Rib{\'o}}, {Richtler}, {Rico},
  {Rieger}, {Riquelme}, {Rivoire}, {Rizi}, {Rodriguez}, {Rodriguez Fernandez},
  {Rodr{\'\i}guez V{\'a}zquez}, {Rojas}, {Romano}, {Romeo}, {Rosado}, {Rovero},
  {Rowell}, {Rudak}, {Rugliancich}, {Rulten}, {Sadeh}, {Safi-Harb}, {Saito},
  {Sakaki}, {Sakurai}, {Salina}, {S{\'a}nchez-Conde}, {Sandaker}, {Sandoval},
  {Sangiorgi}, {Sanguillon}, {Sano}, {Santander}, {Sarkar}, {Satalecka},
  {Saturni}, {Schioppa}, {Schlenstedt}, {Schneider}, {Schoorlemmer},
  {Schovanek}, {Schulz}, {Schussler}, {Schwanke}, {Sciacca}, {Scuderi},
  {Seitenzahl}, {Semikoz}, {Sergijenko}, {Servillat}, {Shalchi}, {Shellard},
  {Sidoli}, {Siejkowski}, {Sillanp{\"a}{\"a}}, {Sironi}, {Sitarek}, {Sliusar},
  {Slowikowska}, {Sol}, {Stamerra}, {Stani{\v{c}}}, {Starling}, {Stawarz},
  {Stefanik}, {Stephan}, {Stolarczyk}, {Stratta}, {Straumann}, {Suomijarvi},
  {Supanitsky}, {Tagliaferri}, {Tajima}, {Tavani}, {Tavecchio}, {Tavernet},
  {Tayabaly}, {Tejedor}, {Temnikov}, {Terada}, {Terrier}, {Terzic}, {Teshima},
  {Testa}, {Thoudam}, {Tian}, {Tibaldo}, {Tluczykont}, {Todero Peixoto},
  {Tokanai}, {Tomastik}, {Tonev}, {Tornikoski}, {Torres}, {Torresi}, {Tosti},
  {Tothill}, {Tovmassian}, {Travnicek}, {Trichard}, {Trifoglio}, {Troyano
  Pujadas}, {Tsujimoto}, {Umana}, {Vagelli}, {Vagnetti}, {Valentino},
  {Vallania}, {Valore}, {van Eldik}, {Vandenbroucke}, {Varner}, {Vasileiadis},
  {Vassiliev}, {V{\'a}zquez Acosta}, {Vecchi}, {Vega}, {Vercellone}, {Veres},
  {Vergani}, {Verzi}, {Vettolani}, {Viana}, {Vigorito}, {Villanueva}, {Voelk},
  {Vollhardt}, {Vorobiov}, {Vrastil}, {Vuillaume}, {Wagner}, {Wagner},
  {Walter}, {Ward}, {Warren}, {Watson}, {Werner}, {White}, {White},
  {Wierzcholska}, {Wilcox}, {Will}, {Williams}, {Wischnewski}, {Wood},
  {Yamamoto}, {Yamazaki}, {Yanagita}, {Yang}, {Yoshida}, {Yoshiike},
  {Yoshikoshi}, {Zacharias}, {Zaharijas}, {Zampieri}, {Zandanel}, {Zanin},
  {Zavrtanik}, {Zavrtanik}, {Zdziarski}, {Zech}, {Zechlin}, {Zhdanov},
  {Ziegler}, \& {Zorn}}]{CTABOOK2019}
{Cherenkov Telescope Array Consortium}, {Acharya}, B.~S., {Agudo}, I., {et~al.}
  2019, {Science with the Cherenkov Telescope Array}

\bibitem[{{Condon}(1992)}]{Condon1992}
{Condon}, J.~J. 1992, \araa, 30, 575

\bibitem[{{Cortese} {et~al.}(2012){Cortese}, {Boissier}, {Boselli}, {Bendo},
  {Buat}, {Davies}, {Eales}, {Heinis}, {Isaak}, \& {Madden}}]{Cortese2012}
{Cortese}, L., {Boissier}, S., {Boselli}, A., {et~al.} 2012, \aap, 544, A101

\bibitem[{{da Cunha} {et~al.}(2008){da Cunha}, {Charlot}, \&
  {Elbaz}}]{Cunha2008}
{da Cunha}, E., {Charlot}, S., \& {Elbaz}, D. 2008, \mnras, 388, 1595

\bibitem[{{Do} {et~al.}(2021){Do}, {Duong}, {McDaniel}, {O'Connor}, {Profumo},
  {Rafael}, {Sweeney}, \& {Vera}}]{Do2021}
{Do}, A., {Duong}, M., {McDaniel}, A., {et~al.} 2021, \prd, 104, 123016

\bibitem[{{Domingo-Santamar{\'\i}a} \& {Torres}(2005)}]{Domingo2005}
{Domingo-Santamar{\'\i}a}, E. \& {Torres}, D.~F. 2005, \aap, 444, 403

\bibitem[{{Dom{\'\i}nguez} {et~al.}(2011){Dom{\'\i}nguez}, {Primack},
  {Rosario}, {Prada}, {Gilmore}, {Faber}, {Koo}, {Somerville},
  {P{\'e}rez-Torres}, {P{\'e}rez-Gonz{\'a}lez}, {Huang}, {Davis},
  {Guhathakurta}, {Barmby}, {Conselice}, {Lozano}, {Newman}, \&
  {Cooper}}]{Dominguez2011}
{Dom{\'\i}nguez}, A., {Primack}, J.~R., {Rosario}, D.~J., {et~al.} 2011,
  \mnras, 410, 2556

\bibitem[{{Donath} {et~al.}(2023){Donath}, {Terrier}, {Remy}, {Sinha}, {Nigro},
  {Pintore}, {Kh{\'e}lifi}, {Olivera-Nieto}, {Ruiz}, {Br{\"u}gge}, {Linhoff},
  {Contreras}, {Acero}, {Aguasca-Cabot}, {Berge}, {Bhattacharjee}, {Buchner},
  {Boisson}, {Carreto Fidalgo}, {Chen}, {de Bony de Lavergne}, {de Miranda
  Cardoso}, {Deil}, {F{\"u}{\ss}ling}, {Funk}, {Giunti}, {Hinton}, {Jouvin},
  {King}, {Lefaucheur}, {Lemoine-Goumard}, {Lenain}, {L{\'o}pez-Coto},
  {Mohrmann}, {Morcuende}, {Panny}, {Regeard}, {Saha}, {Siejkowski},
  {Siemiginowska}, {Sip{\H{o}}cz}, {Unbehaun}, {van Eldik}, {Vuillaume}, \&
  {Zanin}}]{Donath2023}
{Donath}, A., {Terrier}, R., {Remy}, Q., {et~al.} 2023, \aap, 678, A157

\bibitem[{{Dorfi} \& {Breitschwerdt}(2012)}]{Dorfi2012}
{Dorfi}, E.~A. \& {Breitschwerdt}, D. 2012, \aap, 540, A77

\bibitem[{{Ducoin} {et~al.}(2020){Ducoin}, {Corre}, {Leroy}, \& {Le
  Floch}}]{2020MNRAS.492.4768D}
{Ducoin}, J.~G., {Corre}, D., {Leroy}, N., \& {Le Floch}, E. 2020, \mnras, 492,
  4768

\bibitem[{{Eckner} {et~al.}(2018){Eckner}, {Hou}, {Serpico}, {Winter},
  {Zaharijas}, {Martin}, {di Mauro}, {Mirabal}, {Petrovic}, {Prodanovic}, \&
  {Vandenbroucke}}]{Eckner2018}
{Eckner}, C., {Hou}, X., {Serpico}, P.~D., {et~al.} 2018, \apj, 862, 79

\bibitem[{{Foschini} {et~al.}(2022){Foschini}, {Lister}, {Andernach}, {Ciroi},
  {Marziani}, {Ant{\'o}n}, {Berton}, {Dalla Bont{\`a}}, {J{\"a}rvel{\"a}},
  {March{\~a}}, {Romano}, {Tornikoski}, {Vercellone}, \&
  {Vietri}}]{2022Univ....8..587F}
{Foschini}, L., {Lister}, M.~L., {Andernach}, H., {et~al.} 2022, Universe, 8,
  587

\bibitem[{{Gavazzi} {et~al.}(2011){Gavazzi}, {Savorgnan}, \&
  {Fumagalli}}]{Gavazzi2011}
{Gavazzi}, G., {Savorgnan}, G., \& {Fumagalli}, M. 2011, \aap, 534, A31

\bibitem[{{Gil de Paz} {et~al.}(2007){Gil de Paz}, {Boissier}, {Madore},
  {Seibert}, {Joe}, {Boselli}, {Wyder}, {Thilker}, {Bianchi}, {Rey}, {Rich},
  {Barlow}, {Conrow}, {Forster}, {Friedman}, {Martin}, {Morrissey}, {Neff},
  {Schiminovich}, {Small}, {Donas}, {Heckman}, {Lee}, {Milliard}, {Szalay}, \&
  {Yi}}]{GildePaz2007}
{Gil de Paz}, A., {Boissier}, S., {Madore}, B.~F., {et~al.} 2007, \apjs, 173,
  185

\bibitem[{{H.E.S.S. Collaboration} {et~al.}(2018){H.E.S.S. Collaboration},
  {Abdalla}, {Aharonian}, {Ait Benkhali}, {Ang{\"u}ner}, {Arakawa}, {Arcaro},
  {Armand}, {Arrieta}, {Backes}, {Barnard}, {Becherini}, {Becker Tjus},
  {Berge}, {Bernhard}, {Bernl{\"o}hr}, {Blackwell}, {B{\"o}ttcher}, {Boisson},
  {Bolmont}, {Bonnefoy}, {Bordas}, {Bregeon}, {Brun}, {Brun}, {Bryan},
  {B{\"u}chele}, {Bulik}, {Bylund}, {Capasso}, {Caroff}, {Carosi}, {Casanova},
  {Cerruti}, {Chakraborty}, {Chandra}, {Chaves}, {Chen}, {Colafrancesco},
  {Condon}, {Davids}, {Deil}, {Devin}, {deWilt}, {Dirson},
  {Djannati-Ata{\"\i}}, {Dmytriiev}, {Donath}, {Drury}, {Dyks}, {Egberts},
  {Emery}, {Ernenwein}, {Eschbach}, {Fegan}, {Fiasson}, {Fontaine}, {Funk},
  {F{\"u}{\ss}ling}, {Gabici}, {Gallant}, {Garrigoux}, {Gat{\'e}}, {Giavitto},
  {Glawion}, {Glicenstein}, {Gottschall}, {Grondin}, {Hahn}, {Haupt},
  {Heinzelmann}, {Henri}, {Hermann}, {Hinton}, {Hofmann}, {Hoischen}, {Holch},
  {Holler}, {Horns}, {Huber}, {Iwasaki}, {Jacholkowska}, {Jamrozy},
  {Jankowsky}, {Jankowsky}, {Jouvin}, {Jung-Richardt}, {Kastendieck},
  {Kat{\textasciiacute}nski}, {Katsuragawa}, {Katz}, {Kerszberg}, {Khangulyan},
  {Kh{\'e}lifi}, {King}, {Klepser}, {K{\textasciiacute}zniak}, {Komin},
  {Kosack}, {Krakau}, {Kraus}, {Kr{\"u}ger}, {Lamanna}, {Lau}, {Lefaucheur},
  {Lemi{\`e}re}, {Lemoine-Goumard}, {Lenain}, {Leser}, {Lohse}, {Lorentz},
  {L{\'o}pez-Coto}, {Lypova}, {Malyshev}, {Marandon}, {Marcowith}, {Mariaud},
  {Mart{\'\i}-Devesa}, {Marx}, {Maurin}, {Meintjes}, {Mitchell}, {Moderski},
  {Mohamed}, {Mohrmann}, {Moulin}, {Murach}, {Nakashima}, {de Naurois},
  {Ndiyavala}, {Niederwanger}, {Niemiec}, {Oakes}, {O'Brien}, {Odaka}, {Ohm},
  {Ostrowski}, {Oya}, {Padovani}, {Panter}, {Parsons}, {Perennes}, {Petrucci},
  {Peyaud}, {Piel}, {Pita}, {Poireau}, {Priyana Noel}, {Prokhorov}, {Prokoph},
  {P{\"u}hlhofer}, {Punch}, {Quirrenbach}, {Raab}, {Rauth}, {Reimer}, {Reimer},
  {Renaud}, {Rieger}, {Rinchiuso}, {Romoli}, {Rowell}, {Rudak}, {Ruiz-Velasco},
  {Sahakian}, {Saito}, {Sanchez}, {Santangelo}, {Sasaki}, {Schlickeiser},
  {Sch{\"u}ssler}, {Schulz}, {Schwanke}, {Schwemmer}, {Seglar-Arroyo},
  {Senniappan}, {Seyffert}, {Shafi}, {Shilon}, {Shiningayamwe}, {Simoni},
  {Sinha}, {Sol}, {Spanier}, {Specovius}, {Spir-Jacob}, {Stawarz}, {Steenkamp},
  {Stegmann}, {Steppa}, {Sushch}, {Takahashi}, {Tavernet}, {Tavernier},
  {Taylor}, {Terrier}, {Tibaldo}, {Tiziani}, {Tluczykont}, {Trichard},
  {Tsirou}, {Tsuji}, {Tuffs}, {Uchiyama}, {van der Walt}, {van Eldik}, {van
  Rensburg}, {van Soelen}, {Vasileiadis}, {Veh}, {Venter}, {Viana}, {Vincent},
  {Vink}, {Voisin}, {V{\"o}lk}, {Vuillaume}, {Wadiasingh}, {Wagner}, {Wagner},
  {Wagner}, {White}, {Wierzcholska}, {W{\"o}rnlein}, {Yang}, {Zaborov},
  {Zacharias}, {Zanin}, {Zdziarski}, {Zech}, {Zefi}, {Ziegler}, {Zorn}, \&
  {{\.Z}ywucka}}]{HESS2018}
{H.E.S.S. Collaboration}, {Abdalla}, H., {Aharonian}, F., {et~al.} 2018, \aap,
  617, A73

\bibitem[{{Hofmann} \& {Zanin}(2023)}]{CTAO2023}
{Hofmann}, W. \& {Zanin}, R. 2023, arXiv e-prints, arXiv:2305.12888

\bibitem[{{IceCube Collaboration} {et~al.}(2022){IceCube Collaboration},
  {Abbasi}, {Ackermann}, {Adams}, {Aguilar}, {Ahlers}, {Ahrens}, {Alameddine},
  {Alispach}, {Alves}, {Amin}, {Andeen}, {Anderson}, {Anton}, {Arg{\"u}elles},
  {Ashida}, {Axani}, {Bai}, {Balagopal}, {Barbano}, {Barwick}, {Bastian},
  {Basu}, {Baur}, {Bay}, {Beatty}, {Becker}, {Becker Tjus}, {Bellenghi},
  {Benzvi}, {Berley}, {Bernardini}, {Besson}, {Binder}, {Bindig}, {Blaufuss},
  {Blot}, {Boddenberg}, {Bontempo}, {Borowka}, {B{\"o}ser}, {Botner},
  {B{\"o}ttcher}, {Bourbeau}, {Bradascio}, {Braun}, {Brinson}, {Bron},
  {Brostean-Kaiser}, {Browne}, {Burgman}, {Burley}, {Busse}, {Campana},
  {Carnie-Bronca}, {Chen}, {Chen}, {Chirkin}, {Choi}, {Clark}, {Clark},
  {Classen}, {Coleman}, {Collin}, {Conrad}, {Coppin}, {Correa}, {Cowen},
  {Cross}, {Dappen}, {Dave}, {de Clercq}, {Delaunay}, {Delgado L{\'o}pez},
  {Dembinski}, {Deoskar}, {Desai}, {Desiati}, {de Vries}, {de Wasseige}, {de
  With}, {Deyoung}, {Diaz}, {D{\'\i}az-V{\'e}lez}, {Dittmer}, {Dujmovic},
  {Dunkman}, {Duvernois}, {Dvorak}, {Ehrhardt}, {Eller}, {Engel}, {Erpenbeck},
  {Evans}, {Evenson}, {Fan}, {Fazely}, {Fedynitch}, {Feigl}, {Fiedlschuster},
  {Fienberg}, {Filimonov}, {Finley}, {Fischer}, {Fox}, {Franckowiak},
  {Friedman}, {Fritz}, {F{\"u}rst}, {Gaisser}, {Gallagher}, {Ganster},
  {Garcia}, {Garrappa}, {Gerhardt}, {Ghadimi}, {Glaser}, {Glauch},
  {Gl{\"u}senkamp}, {Goldschmidt}, {Gonzalez}, {Goswami}, {Grant},
  {Gr{\'e}goire}, {Griswold}, {G{\"u}nther}, {Gutjahr}, {Haack}, {Hallgren},
  {Halliday}, {Halve}, {Halzen}, {Hanson}, {Hardin}, {Harnisch}, {Haungs},
  {Hebecker}, {Helbing}, {Henningsen}, {Hettinger}, {Hickford}, {Hignight},
  {Hill}, {Hill}, {Hoffman}, {Hoffmann}, {Hokanson-Fasig}, {Hoshina}, {Huang},
  {Huber}, {Huber}, {Hultqvist}, {H{\"u}nnefeld}, {Hussain}, {Hymon}, {in},
  {Iovine}, {Ishihara}, {Jansson}, {Japaridze}, {Jeong}, {Jin}, {Jones},
  {Kang}, {Kang}, {Kang}, {Kappes}, {Kappesser}, {Kardum}, {Karg}, {Karl},
  {Karle}, {Katz}, {Kauer}, {Kellermann}, {Kelley}, {Kheirandish}, {Kin},
  {Kintscher}, {Kiryluk}, {Klein}, {Koirala}, {Kolanoski}, {Kontrimas},
  {K{\"o}pke}, {Kopper}, {Kopper}, {Koskinen}, {Koundal}, {Kovacevich},
  {Kowalski}, {Kozynets}, {Kun}, {Kurahashi}, {Lad}, {Lagunas Gualda},
  {Lanfranchi}, {Larson}, {Lauber}, {Lazar}, {Lee}, {Leonard},
  {Leszczy{\'n}ska}, {Li}, {Lincetto}, {Liu}, {Liubarska}, {Lohfink}, {Lozano
  Mariscal}, {Lu}, {Lucarelli}, {Ludwig}, {Luszczak}, {Lyu}, {Ma}, {Madsen},
  {Mahn}, {Makino}, {Mancina}, {Mari{\c{s}}}, {Martinez-Soler}, {Maruyama},
  {Mase}, {McElroy}, {McNally}, {Mead}, {Meagher}, {Mechbal}, {Medina},
  {Meier}, {Meighen-Berger}, {Micallef}, {Mockler}, {Montaruli}, {Moore},
  {Morse}, {Moulai}, {Naab}, {Nagai}, {Nahnhauer}, {Naumann}, {Necker},
  {Nguyen}, {Niederhausen}, {Nisa}, {Nowicki}, {Nygren}, {Obertack},
  {Pollmann}, {Oehler}, {Oeyen}, {Olivas}, {O'Sullivan}, {Pandya}, {Pankova},
  {Park}, {Parker}, {Paudel}, {Paul}, {P{\'e}rez de Los Heros}, {Peters},
  {Peterson}, {Philippen}, {Pieper}, {Pittermann}, {Pizzuto}, {Plum},
  {Popovych}, {Porcelli}, {Prado Rodriguez}, {Price}, {Pries}, {Przybylski},
  {Rack-Helleis}, {Raissi}, {Rameez}, {Rawlins}, {Rea}, {Rehman},
  {Reichherzer}, {Reimann}, {Renzi}, {Resconi}, {Reusch}, {Rhode}, {Richman},
  {Riedel}, {Roberts}, {Robertson}, {Roellinghoff}, {Rongen}, {Rott}, {Ruhe},
  {Ryckbosch}, {Rysewyk Cantu}, {Safa}, {Saffer}, {Sanchez Herrera},
  {Sandrock}, {Sandroos}, {Santander}, {Sarkar}, {Sarkar}, {Satalecka},
  {Schaufel}, {Schieler}, {Schindler}, {Schmidt}, {Schneider}, {Schneider},
  {Schr{\"o}der}, {Schumacher}, {Schwefer}, {Sclafani}, {Seckel}, {Seunarine},
  {Sharma}, {Shefali}, {Silva}, {Skrzypek}, {Smithers}, {Snihur},
  {Soedingrekso}, {Soldin}, {Spannfellner}, {Spiczak}, {Spiering},
  {Stachurska}, {Stamatikos}, {Stanev}, {Stein}, {Stettner}, {Steuer},
  {Stezelberger}, {Stokstad}, {St{\"u}rwald}, {Stuttard}, {Sullivan},
  {Taboada}, {Ter-Antonyan}, {Tilav}, {Tischbein}, {Tollefson}, {T{\"o}nnis},
  {Toscano}, {Tosi}, {Trettin}, {Tselengidou}, {Tung}, {Turcati}, {Turcotte},
  {Turley}, {Twagirayezu}, {Ty}, {Unland Elorrieta}, {Valtonen-Mattila},
  {Vandenbroucke}, {van Eijndhoven}, {Vannerom}, {van Santen}, {Verpoest},
  {Walck}, {Watson}, {Weaver}, {Weigel}, {Weindl}, {Weiss}, {Weldert}, {Wendt},
  {Werthebach}, {Weyrauch}, {Whitehorn}, {Wiebusch}, {Williams}, {Wolf},
  {Woschnagg}, {Wrede}, {Wulff}, {Xu}, {Yanez}, {Yoshida}, {Yu}, {Yuan},
  {Zhangan}, \& {Zhelnin}}]{IceCube2022}
{IceCube Collaboration}, {Abbasi}, R., {Ackermann}, M., {et~al.} 2022, Science,
  378, 538

\bibitem[{{Karachentsev} {et~al.}(2018){Karachentsev}, {Kaisina}, \&
  {Makarov}}]{2018MNRAS.479.4136K}
{Karachentsev}, I.~D., {Kaisina}, E.~I., \& {Makarov}, D.~I. 2018, \mnras, 479,
  4136

\bibitem[{{Kaur} {et~al.}(2019){Kaur}, {Falcone}, {Stroh}, {Kennea}, \&
  {Ferrara}}]{2019ApJ...887...18K}
{Kaur}, A., {Falcone}, A.~D., {Stroh}, M.~D., {Kennea}, J.~A., \& {Ferrara},
  E.~C. 2019, \apj, 887, 18

\bibitem[{{Kennicutt}(1998)}]{kennikutt1998A}
{Kennicutt}, Robert~C., J. 1998, \apj, 498, 541

\bibitem[{{Kennicutt} {et~al.}(2008){Kennicutt}, {Lee}, {Funes}, {J.}, {Sakai},
  \& {Akiyama}}]{Kennicutt2008}
{Kennicutt}, Robert~C., J., {Lee}, J.~C., {Funes}, J.~G., {et~al.} 2008, \apjs,
  178, 247

\bibitem[{{Kennicutt} \& {Evans}(2012)}]{Kennicutt2012}
{Kennicutt}, R.~C. \& {Evans}, N.~J. 2012, \araa, 50, 531

\bibitem[{{Kornecki} {et~al.}(2020){Kornecki}, {Pellizza}, {del Palacio},
  {M{\"u}ller}, {Albacete-Colombo}, \& {Romero}}]{K20}
{Kornecki}, P., {Pellizza}, L.~J., {del Palacio}, S., {et~al.} 2020, \aap, 641,
  A147

\bibitem[{{Kornecki} {et~al.}(2022){Kornecki}, {Peretti}, {del Palacio},
  {Benaglia}, \& {Pellizza}}]{K22}
{Kornecki}, P., {Peretti}, E., {del Palacio}, S., {Benaglia}, P., \&
  {Pellizza}, L.~J. 2022, \aap, 657, A49

\bibitem[{Kornecki {et~al.}(2023)Kornecki, Peretti, del Palacio, Marcowith, \&
  Araudo}]{Kornecki:2023Ia}
Kornecki, P., Peretti, E., del Palacio, S., Marcowith, A., \& Araudo, A. 2023,
  PoS, Gamma2022, 216

\bibitem[{{Krumholz} {et~al.}(2020){Krumholz}, {Crocker}, {Xu}, {Lazarian},
  {Rosevear}, \& {Bedwell-Wilson}}]{Krumholz2020}
{Krumholz}, M.~R., {Crocker}, R.~M., {Xu}, S., {et~al.} 2020, \mnras, 493, 2817

\bibitem[{{Lacki} \& {Thompson}(2013)}]{Lacki2013}
{Lacki}, B.~C. \& {Thompson}, T.~A. 2013, \apj, 762, 29

\bibitem[{{Lacki} {et~al.}(2011){Lacki}, {Thompson}, {Quataert}, {Loeb}, \&
  {Waxman}}]{Lacki2011}
{Lacki}, B.~C., {Thompson}, T.~A., {Quataert}, E., {Loeb}, A., \& {Waxman}, E.
  2011, \apj, 734, 107

\bibitem[{{Lamastra} {et~al.}(2019){Lamastra}, {Tavecchio}, {Romano},
  {Landoni}, \& {Vercellone}}]{Lamastra2019}
{Lamastra}, A., {Tavecchio}, F., {Romano}, P., {Landoni}, M., \& {Vercellone},
  S. 2019, Astroparticle Physics, 112, 16

\bibitem[{{Lenain} {et~al.}(2010){Lenain}, {Ricci}, {T{\"u}rler}, {Dorner}, \&
  {Walter}}]{Lenain2010}
{Lenain}, J.~P., {Ricci}, C., {T{\"u}rler}, M., {Dorner}, D., \& {Walter}, R.
  2010, \aap, 524, A72

\bibitem[{{Lenc} \& {Tingay}(2009)}]{Lenc2009}
{Lenc}, E. \& {Tingay}, S.~J. 2009, \aj, 137, 537

\bibitem[{{LHAASO collaboration}(2021)}]{LHAASO_sens2021}
{LHAASO collaboration}. 2021, arXiv e-prints, arXiv:2101.03508

\bibitem[{{Mannheim} {et~al.}(2012){Mannheim}, {Els{\"a}sser}, \&
  {Tibolla}}]{Mannheim2012}
{Mannheim}, K., {Els{\"a}sser}, D., \& {Tibolla}, O. 2012, Astroparticle
  Physics, 35, 797

\bibitem[{{Martin}(2014)}]{Martin2014}
{Martin}, P. 2014, \aap, 564, A61

\bibitem[{{McDaniel} {et~al.}(2019){McDaniel}, {Jeltema}, \&
  {Profumo}}]{McDaniel2019}
{McDaniel}, A., {Jeltema}, T., \& {Profumo}, S. 2019, \prd, 100, 023014

\bibitem[{{Nasonova} {et~al.}(2011){Nasonova}, {de Freitas Pacheco}, \&
  {Karachentsev}}]{2011A&A...532A.104N}
{Nasonova}, O.~G., {de Freitas Pacheco}, J.~A., \& {Karachentsev}, I.~D. 2011,
  \aap, 532, A104

\bibitem[{Observatory \& Consortium(2021)}]{CTAsens}
Observatory, C. T.~A. \& Consortium, C. T.~A. 2021, {CTAO Instrument Response
  Functions - prod5 version v0.1}

\bibitem[{{Ohm} \& {Hinton}(2013)}]{Ohm2013}
{Ohm}, S. \& {Hinton}, J.~A. 2013, \mnras, 429, L70

\bibitem[{{Peretti} {et~al.}(2019){Peretti}, {Blasi}, {Aharonian}, \&
  {Morlino}}]{Peretti2019}
{Peretti}, E., {Blasi}, P., {Aharonian}, F., \& {Morlino}, G. 2019, \mnras,
  487, 168

\bibitem[{{Peretti} {et~al.}(2020){Peretti}, {Blasi}, {Aharonian}, {Morlino},
  \& {Cristofari}}]{Peretti2020}
{Peretti}, E., {Blasi}, P., {Aharonian}, F., {Morlino}, G., \& {Cristofari}, P.
  2020, \mnras, 493, 5880

\bibitem[{{Peretti} {et~al.}(2023){Peretti}, {Lamastra}, {Saturni}, {Ahlers},
  {Blasi}, {Morlino}, \& {Cristofari}}]{Peretti2023}
{Peretti}, E., {Lamastra}, A., {Saturni}, F.~G., {et~al.} 2023, \mnras, 526,
  181

\bibitem[{{Peretti} {et~al.}(2022){Peretti}, {Morlino}, {Blasi}, \&
  {Cristofari}}]{Peretti2022}
{Peretti}, E., {Morlino}, G., {Blasi}, P., \& {Cristofari}, P. 2022, \mnras,
  511, 1336

\bibitem[{{P{\'e}rez-Beaupuits} {et~al.}(2011){P{\'e}rez-Beaupuits}, {Spoon},
  {Spaans}, \& {Smith}}]{Perez2011}
{P{\'e}rez-Beaupuits}, J.~P., {Spoon}, H.~W.~W., {Spaans}, M., \& {Smith},
  J.~D. 2011, \aap, 533, A56

\bibitem[{{Persic} {et~al.}(2008){Persic}, {Rephaeli}, \&
  {Arieli}}]{Persic2008}
{Persic}, M., {Rephaeli}, Y., \& {Arieli}, Y. 2008, \aap, 486, 143

\bibitem[{{Persic} {et~al.}(2024){Persic}, {Rephaeli}, \& {Rando}}]{Persic2024}
{Persic}, M., {Rephaeli}, Y., \& {Rando}, R. 2024, \aap, 685, A47

\bibitem[{{Pfrommer} {et~al.}(2017){Pfrommer}, {Pakmor}, {Simpson}, \&
  {Springel}}]{Pfrommer2017}
{Pfrommer}, C., {Pakmor}, R., {Simpson}, C.~M., \& {Springel}, V. 2017, \apjl,
  847, L13

\bibitem[{{Planck Collaboration} {et~al.}(2020){Planck Collaboration},
  {Aghanim}, {Akrami}, {Ashdown}, {Aumont}, {Baccigalupi}, {Ballardini},
  {Banday}, {Barreiro}, {Bartolo}, {Basak}, {Battye}, {Benabed}, {Bernard},
  {Bersanelli}, {Bielewicz}, {Bock}, {Bond}, {Borrill}, {Bouchet}, {Boulanger},
  {Bucher}, {Burigana}, {Butler}, {Calabrese}, {Cardoso}, {Carron},
  {Challinor}, {Chiang}, {Chluba}, {Colombo}, {Combet}, {Contreras}, {Crill},
  {Cuttaia}, {de Bernardis}, {de Zotti}, {Delabrouille}, {Delouis}, {Di
  Valentino}, {Diego}, {Dor{\'e}}, {Douspis}, {Ducout}, {Dupac}, {Dusini},
  {Efstathiou}, {Elsner}, {En{\ss}lin}, {Eriksen}, {Fantaye}, {Farhang},
  {Fergusson}, {Fernandez-Cobos}, {Finelli}, {Forastieri}, {Frailis},
  {Fraisse}, {Franceschi}, {Frolov}, {Galeotta}, {Galli}, {Ganga},
  {G{\'e}nova-Santos}, {Gerbino}, {Ghosh}, {Gonz{\'a}lez-Nuevo}, {G{\'o}rski},
  {Gratton}, {Gruppuso}, {Gudmundsson}, {Hamann}, {Handley}, {Hansen},
  {Herranz}, {Hildebrandt}, {Hivon}, {Huang}, {Jaffe}, {Jones}, {Karakci},
  {Keih{\"a}nen}, {Keskitalo}, {Kiiveri}, {Kim}, {Kisner}, {Knox},
  {Krachmalnicoff}, {Kunz}, {Kurki-Suonio}, {Lagache}, {Lamarre}, {Lasenby},
  {Lattanzi}, {Lawrence}, {Le Jeune}, {Lemos}, {Lesgourgues}, {Levrier},
  {Lewis}, {Liguori}, {Lilje}, {Lilley}, {Lindholm}, {L{\'o}pez-Caniego},
  {Lubin}, {Ma}, {Mac{\'\i}as-P{\'e}rez}, {Maggio}, {Maino}, {Mandolesi},
  {Mangilli}, {Marcos-Caballero}, {Maris}, {Martin}, {Martinelli},
  {Mart{\'\i}nez-Gonz{\'a}lez}, {Matarrese}, {Mauri}, {McEwen}, {Meinhold},
  {Melchiorri}, {Mennella}, {Migliaccio}, {Millea}, {Mitra},
  {Miville-Desch{\^e}nes}, {Molinari}, {Montier}, {Morgante}, {Moss}, {Natoli},
  {N{\o}rgaard-Nielsen}, {Pagano}, {Paoletti}, {Partridge}, {Patanchon},
  {Peiris}, {Perrotta}, {Pettorino}, {Piacentini}, {Polastri}, {Polenta},
  {Puget}, {Rachen}, {Reinecke}, {Remazeilles}, {Renzi}, {Rocha}, {Rosset},
  {Roudier}, {Rubi{\~n}o-Mart{\'\i}n}, {Ruiz-Granados}, {Salvati}, {Sandri},
  {Savelainen}, {Scott}, {Shellard}, {Sirignano}, {Sirri}, {Spencer},
  {Sunyaev}, {Suur-Uski}, {Tauber}, {Tavagnacco}, {Tenti}, {Toffolatti},
  {Tomasi}, {Trombetti}, {Valenziano}, {Valiviita}, {Van Tent}, {Vibert},
  {Vielva}, {Villa}, {Vittorio}, {Wandelt}, {Wehus}, {White}, {White},
  {Zacchei}, \& {Zonca}}]{plank2020}
{Planck Collaboration}, {Aghanim}, N., {Akrami}, Y., {et~al.} 2020, \aap, 641,
  A6

\bibitem[{{Prieto} {et~al.}(2004){Prieto}, {Meisenheimer}, {Marco}, {Reunanen},
  {Contini}, {Clenet}, {Davies}, {Gratadour}, {Henning}, {Klaas}, {Kotilainen},
  {Leinert}, {Lutz}, {Rouan}, \& {Thatte}}]{Prieto2004}
{Prieto}, M.~A., {Meisenheimer}, K., {Marco}, O., {et~al.} 2004, \apj, 614, 135

\bibitem[{{Rice} {et~al.}(1988){Rice}, {Lonsdale}, {Soifer}, {Neugebauer},
  {Kopan}, {Lloyd}, {de Jong}, \& {Habing}}]{rice1988}
{Rice}, W., {Lonsdale}, C.~J., {Soifer}, B.~T., {et~al.} 1988, \apjs, 68, 91

\bibitem[{{Rojas-Bravo} \& {Araya}(2016)}]{RojasBravo2016}
{Rojas-Bravo}, C. \& {Araya}, M. 2016, \mnras, 463, 1068

\bibitem[{{Romero} {et~al.}(2018){Romero}, {M{\"u}ller}, \&
  {Roth}}]{Romero2018}
{Romero}, G.~E., {M{\"u}ller}, A.~L., \& {Roth}, M. 2018, \aap, 616, A57

\bibitem[{{Roth} {et~al.}(2021){Roth}, {Krumholz}, {Crocker}, \&
  {Celli}}]{Roth2021}
{Roth}, M.~A., {Krumholz}, M.~R., {Crocker}, R.~M., \& {Celli}, S. 2021, \nat,
  597, 341

\bibitem[{{Roth} {et~al.}(2023){Roth}, {Krumholz}, {Crocker}, \&
  {Thompson}}]{Roth2023}
{Roth}, M.~A., {Krumholz}, M.~R., {Crocker}, R.~M., \& {Thompson}, T.~A. 2023,
  \mnras, 523, 2608

\bibitem[{{Sanders} {et~al.}(2003){Sanders}, {Mazzarella}, {Kim}, {Surace}, \&
  {Soifer}}]{Sanders2003}
{Sanders}, D.~B., {Mazzarella}, J.~M., {Kim}, D.~C., {Surace}, J.~A., \&
  {Soifer}, B.~T. 2003, \aj, 126, 1607

\bibitem[{{Shimono} {et~al.}(2021){Shimono}, {Totani}, \&
  {Sudoh}}]{Shimono2021}
{Shimono}, N., {Totani}, T., \& {Sudoh}, T. 2021, \mnras, 506, 6212

\bibitem[{{Strickland} \& {Heckman}(2009)}]{Strickland2009}
{Strickland}, D.~K. \& {Heckman}, T.~M. 2009, \apj, 697, 2030

\bibitem[{{Strong} {et~al.}(2011){Strong}, {Orlando}, \& {Jaffe}}]{Strong2011}
{Strong}, A.~W., {Orlando}, E., \& {Jaffe}, T.~R. 2011, \aap, 534, A54

\bibitem[{{Sudoh} {et~al.}(2018){Sudoh}, {Totani}, \& {Kawanaka}}]{Sudoh2018}
{Sudoh}, T., {Totani}, T., \& {Kawanaka}, N. 2018, \pasj, 70, 49

\bibitem[{{Tombesi} {et~al.}(2010){Tombesi}, {Cappi}, {Reeves}, {Palumbo},
  {Yaqoob}, {Braito}, \& {Dadina}}]{Tombesi2010}
{Tombesi}, F., {Cappi}, M., {Reeves}, J.~N., {et~al.} 2010, \aap, 521, A57

\bibitem[{{Tully} {et~al.}(2023){Tully}, {Kourkchi}, {Courtois}, {Anand},
  {Blakeslee}, {Brout}, {Jaeger}, {Dupuy}, {Guinet}, {Howlett}, {Jensen},
  {Pomar{\`e}de}, {Rizzi}, {Rubin}, {Said}, {Scolnic}, \& {Stahl}}]{Tully2023}
{Tully}, R.~B., {Kourkchi}, E., {Courtois}, H.~M., {et~al.} 2023, \apj, 944, 94

\bibitem[{{Vecchiotti} {et~al.}(2022){Vecchiotti}, {Pagliaroli}, \&
  {Villante}}]{Vecchiotti2022}
{Vecchiotti}, V., {Pagliaroli}, G., \& {Villante}, F.~L. 2022, Communications
  Physics, 5, 161

\bibitem[{{Veilleux} {et~al.}(2005){Veilleux}, {Cecil}, \&
  {Bland-Hawthorn}}]{Veilleux2005}
{Veilleux}, S., {Cecil}, G., \& {Bland-Hawthorn}, J. 2005, \araa, 43, 769

\bibitem[{{VERITAS Collaboration} {et~al.}(2009){VERITAS Collaboration},
  {Acciari}, {Aliu}, {Arlen}, {Aune}, {Bautista}, {Beilicke}, {Benbow},
  {Boltuch}, {Bradbury}, {Buckley}, {Bugaev}, {Byrum}, {Cannon}, {Celik},
  {Cesarini}, {Chow}, {Ciupik}, {Cogan}, {Colin}, {Cui}, {Dickherber}, {Duke},
  {Fegan}, {Finley}, {Finnegan}, {Fortin}, {Fortson}, {Furniss}, {Galante},
  {Gall}, {Gibbs}, {Gillanders}, {Godambe}, {Grube}, {Guenette}, {Gyuk},
  {Hanna}, {Holder}, {Horan}, {Hui}, {Humensky}, {Imran}, {Kaaret}, {Karlsson},
  {Kertzman}, {Kieda}, {Kildea}, {Konopelko}, {Krawczynski}, {Krennrich},
  {Lang}, {Lebohec}, {Maier}, {McArthur}, {McCann}, {McCutcheon}, {Millis},
  {Moriarty}, {Mukherjee}, {Nagai}, {Ong}, {Otte}, {Pandel}, {Perkins},
  {Pizlo}, {Pohl}, {Quinn}, {Ragan}, {Reyes}, {Reynolds}, {Roache}, {Rose},
  {Schroedter}, {Sembroski}, {Smith}, {Steele}, {Swordy}, {Theiling},
  {Thibadeau}, {Varlotta}, {Vassiliev}, {Vincent}, {Wagner}, {Wakely}, {Ward},
  {Weekes}, {Weinstein}, {Weisgarber}, {Williams}, {Wissel}, {Wood}, \&
  {Zitzer}}]{VERITAS2009}
{VERITAS Collaboration}, {Acciari}, V.~A., {Aliu}, E., {et~al.} 2009, \nat,
  462, 770

\bibitem[{{V{\"o}lk} {et~al.}(1996){V{\"o}lk}, {Aharonian}, \&
  {Breitschwerdt}}]{Volk1996}
{V{\"o}lk}, H.~J., {Aharonian}, F.~A., \& {Breitschwerdt}, D. 1996, \ssr, 75,
  279

\bibitem[{{Wang} \& {Fields}(2018)}]{Wang2018}
{Wang}, X. \& {Fields}, B.~D. 2018, \mnras, 474, 4073

\bibitem[{{Werhahn} {et~al.}(2021){Werhahn}, {Pfrommer}, {Girichidis}, \&
  {Winner}}]{Werhahn2021}
{Werhahn}, M., {Pfrommer}, C., {Girichidis}, P., \& {Winner}, G. 2021, \mnras,
  505, 3295

\bibitem[{{Wojaczy{\'n}ski} \& {Nied{\'z}wiecki}(2017)}]{Wojaczy2017}
{Wojaczy{\'n}ski}, R. \& {Nied{\'z}wiecki}, A. 2017, \apj, 849, 97

\bibitem[{{Wood} {et~al.}(2017){Wood}, {Caputo}, {Charles}, {Di Mauro},
  {Magill}, {Perkins}, \& {Fermi-LAT Collaboration}}]{Wood+2017}
{Wood}, M., {Caputo}, R., {Charles}, E., {et~al.} 2017, in International Cosmic
  Ray Conference, Vol. 301, 35th International Cosmic Ray Conference
  (ICRC2017), 824

\bibitem[{{Wright} {et~al.}(2010){Wright}, {Eisenhardt}, {Mainzer}, {Ressler},
  {Cutri}, {Jarrett}, {Kirkpatrick}, {Padgett}, {McMillan}, {Skrutskie},
  {Stanford}, {Cohen}, {Walker}, {Mather}, {Leisawitz}, {Gautier}, {McLean},
  {Benford}, {Lonsdale}, {Blain}, {Mendez}, {Irace}, {Duval}, {Liu}, {Royer},
  {Heinrichsen}, {Howard}, {Shannon}, {Kendall}, {Walsh}, {Larsen}, {Cardon},
  {Schick}, {Schwalm}, {Abid}, {Fabinsky}, {Naes}, \& {Tsai}}]{WISE2010}
{Wright}, E.~L., {Eisenhardt}, P. R.~M., {Mainzer}, A.~K., {et~al.} 2010, \aj,
  140, 1868

\bibitem[{{Xi} {et~al.}(2020{\natexlab{a}}){Xi}, {Liu}, {Wang}, {Yang}, {Yuan},
  \& {Zhang}}]{Xi2020}
{Xi}, S.-Q., {Liu}, R.-Y., {Wang}, X.-Y., {et~al.} 2020{\natexlab{a}}, \apjl,
  896, L33

\bibitem[{{Xi} {et~al.}(2020{\natexlab{b}}){Xi}, {Zhang}, {Liu}, \&
  {Wang}}]{Xi2020A}
{Xi}, S.-Q., {Zhang}, H.-M., {Liu}, R.-Y., \& {Wang}, X.-Y. 2020{\natexlab{b}},
  \apj, 901, 158

\bibitem[{{Xiang} {et~al.}(2021){Xiang}, {Jiang}, \& {Tang}}]{Xiang2021}
{Xiang}, Y.-C., {Jiang}, Z.-J., \& {Tang}, Y.-Y. 2021, Research in Astronomy
  and Astrophysics, 21, 263

\bibitem[{{Xing} \& {Wang}(2023)}]{Xing2023}
{Xing}, Y. \& {Wang}, Z. 2023, \apj, 952, 112

\bibitem[{{Xing} {et~al.}(2023){Xing}, {Wang}, {Zheng}, \& {Li}}]{Yi2023}
{Xing}, Y., {Wang}, Z., {Zheng}, D., \& {Li}, J. 2023, \apjl, 945, L22

\bibitem[{{Yang} \& {Razzaque}(2019)}]{Yang2019}
{Yang}, L. \& {Razzaque}, S. 2019, \prd, 99, 083007

\bibitem[{{Yoast-Hull} {et~al.}(2013){Yoast-Hull}, {Everett}, {Gallagher}, \&
  {Zweibel}}]{Yoast-Hull2013}
{Yoast-Hull}, T.~M., {Everett}, J.~E., {Gallagher}, J.~S., I., \& {Zweibel},
  E.~G. 2013, \apj, 768, 53

\bibitem[{{Yun} {et~al.}(2001){Yun}, {Reddy}, \& {Condon}}]{Yun2001}
{Yun}, M.~S., {Reddy}, N.~A., \& {Condon}, J.~J. 2001, \apj, 554, 803

\end{thebibliography}
%
\appendix
\onecolumn

\renewcommand{\thetable}{A.\arabic{table}}
\section{Complementary tables and plots}
\label{app: complement}

\begin{figure*}[hb!]
    \centering
    \begin{minipage}[t]{0.45\textwidth} 
        \centering
        \includegraphics[width=\textwidth]{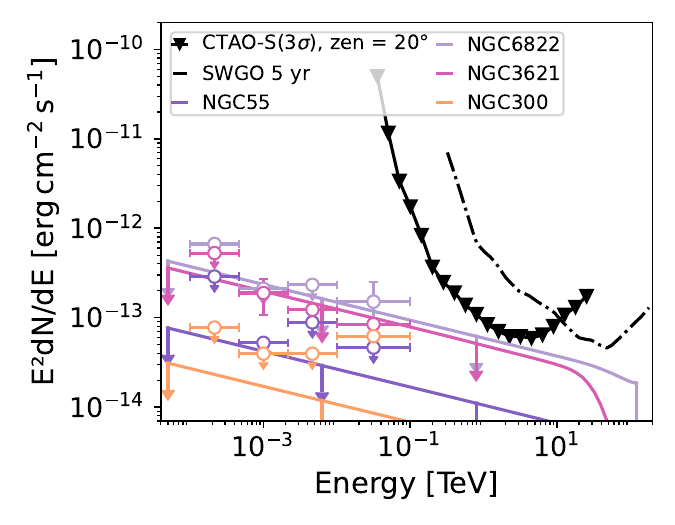}
    \end{minipage}%
    \hspace{0.04\textwidth}
    \begin{minipage}[t]{0.45\textwidth} 
        \centering
        \includegraphics[width=\textwidth]{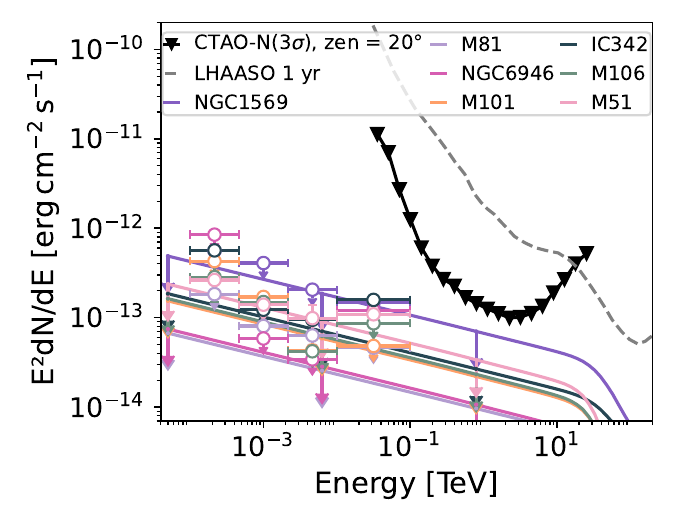}
    \end{minipage}
    \caption{SEDs of non-selected galaxies from the Southern (\textit{left panel}) and Northern (\textit{right panel}) hemispheres. The sensitivity lines shown in black match those in Fig.~\ref{fig: SED_best_non_det}. The spectral points and upper limits are in different colors for each galaxy, as labeled in the figure.
}
    \label{fig: SED_other}
\end{figure*}

In this appendix, we present additional information that complement the findings discussed in the main body of the paper. Table~\ref{tab: excl} lists the \textit{Fermi}-LAT sources catalog in the 4FGL-DR4 that could be associated with SFGs but were not identified as such in multiwavelength investigation. Figure~\ref{fig: SED_other} provides upper limits on the spectra of candidate SFGs that are deemed of secondary interest for VHE observation.

\begin{table}[h!]
\caption{\label{tab: excl} 4FGL-DR4 sources rejected from the sample, as discussed in Sec~\ref{subsec: GeVDet_sel_crit}.}
\begin{tabular}{lrrcll}
\hline \hline 
4FGL Name & R.A. & Dec & CLASS & ASSOC & Exclusion reason \\
\hline
J0003.1-1543\phantom{a} & \phantom{00}0.80 & -15.73 & sey & PKS 0000-160 & Blazar-like broad-band SED \\
J0112.5-0651\phantom{a} & \phantom{0}18.13 & \phantom{0}-6.86 & sey & WISEA J011252.37-064057.1 & O-IR SED not clearly SFG-like \\
J0637.3-6220\phantom{a} & \phantom{0}99.34 & -62.34 & sey & WISEA J063652.55-622035.7 & O-IR SED not clearly SFG-like \\
J0859.8+0053\phantom{a} & 134.95          & \phantom{-0}0.90 & sey & WISEA J085956.49+005244.1 & O-IR SED not clearly SFG-like \\
J1015.1-6353\phantom{a} & 153.78          & -63.88 & sey & WISEA J101433.06-635155.0 & Giant-elliptical-like O-IR SED\\
J1114.2+0638\phantom{a} & 168.56          & \phantom{-0}6.63 & gal & IC 678 & Giant-elliptical-like O-IR SED \\
J1308.9-5730\phantom{a} & 197.23          & -57.50 & sbg & WISEA J130831.60-572649.2 & O-IR SED not clearly SFG-like \\
J1357.3+3730\phantom{a} & 209.33          & \phantom{-}37.50 & gal & NGC\,5380 & Giant-elliptical-like O-IR SED \\
J1438.0+0219\phantom{a} & 219.50          & \phantom{-0}2.33 & sbg & WISEA J143747.60+021733.5 & O-IR SED not clearly SFG-like \\
J1603.6-0451\phantom{a} & 240.92          & \phantom{0}-4.86 & sey & WISEA J160325.44-044907.7 & O-IR SED not clearly SFG-like \\
J1651.1-5848\phantom{a} & 252.78          & -58.81 & sbg & WISEA J165121.14-590011.5 & O-IR SED not clearly SFG-like \\
J1737.1-2901\phantom{a} & 264.28          & -29.03 & sey & WISEA J173737.21-290825.4 & Giant-elliptical-like O-IR SED \\
J1821.6+6636\phantom{a} & 275.41          & \phantom{-}66.61 & sey & WISEA J182223.43+663751.2 & O-IR SED not clearly SFG-like \\
J2118.8-0723c           & 319.72          & \phantom{0}-7.39 & sey & TXS 2116-077 & Blazar-like broad-band SED \\
J2210.4-0930\phantom{a} & 332.60          & \phantom{0}-9.50 & sey & WISEA J221031.71-093158.5 & O-IR SED not clearly SFG-like \\
\hline \hline 
\end{tabular}
\end{table}

\renewcommand{\thetable}{B.\arabic{table}}
\section{Summary table}
\label{app: summary}

Table~\ref{tab: summary} gives an overview of the SFGs studied here and the level of interest for the new $\gamma$-ray observatories. The first four columns provide information about the identification and location of the SFGs. The next three columns indicate the samples for which the SFGs verify the selection criteria defined in section~\ref{sec: nonGeVDet}. It should be noted that the five most distant sources do not satisfy these criteria for any sample, although the galaxies are detected by \textit{Fermi}-LAT. The next three columns give the SFR and the constraints on the $\gamma$-ray flux. The last two columns qualify the level of interest of the observation of each SFG by the CTAO, LHAASO, and SWGO, thus summarizing the discussion provided in section~\ref{sec: disc}.

\begin{landscape}
\begin{table}[h!]
\caption{\label{tab: summary} Sample of SFGs studied in this work.  }
\begin{tabular}{lcrrccccllcc}
\hline \hline 
Name 		&	 $d$ 			&	 R.A. 		&	 Dec		&	Revised		&	IRAS 	&	Ackermann	&	 $\dot{M}_*$ 		&	 4FGL Name 		&	$\mathcal{F}_\mathrm{2\,GeV}$	&	CTAO	&	LHAASO,\\
 		&	 $\mathrm{Mpc}$ 	&	 deg 		&	 deg		&	MANGROVE	&	RBGS	&	  2012 sample		&	 $M_{\odot}\,\mathrm{yr^{-1}}$	&			&	 $10^{-13}\,\mathrm{erg\,s^{-1}\,cm^{-2}}$	&		&	SWGO \\
\hline																							
LMC 		&	 $0.049 \pm 0.002$   	&	 80.00	 	&	 -68.75		&	$\checkmark$	&	$\checkmark$	&	$\checkmark$	&	 $0.192 \pm 0.022$   	&	 J0519.9-6845e	&	 $240.0 \pm 5.9$ 		&	?	&	?	\\
SMC 		&	 $0.064 \pm 0.001$   	&	 14.50	 	&	 -72.75		&	$\checkmark$	&	$\checkmark$	&	$\checkmark$	&	 $0.030 \pm 0.003$   	&	 J0058.0-7245e   	&	 $50.3 \pm 2.5$ 		&	?	&	?	\\
NGC\,6822 	&	 $0.4603 \pm 0.0040$ 	&	296.24		&	-14.80		&	$\checkmark$	&			&			&	 $0.0070 \pm 0.0006$  	&		-		&	${<}\,2.05$			&	-	&	-	\\
M31 		&	 $0.74 \pm 0.02$   	&	 10.82	 	&	 41.24		&	$\checkmark$	&	$\checkmark$	&	$\checkmark$	&	 $0.245 \pm 0.023$  	&	 J0043.2+4114    	&	 $3.91 \pm 0.55$ 		&	?	&	-	\\
M33     	&	 $0.853 \pm 0.016$   	&	 23.46 		&	 30.66 		&	$\checkmark$	&	$\checkmark$	&	$\checkmark$	&	 $0.251 \pm 0.013$  	&		-		&	${<}\,2.49$			&	+	&	-	\\
NGC\,300 	&	 $1.944 \pm 0.040$  	&	 13.72 		&	-37.68		&	$\checkmark$	&			&			&	 $0.138 \pm 0.005$  	&		-		&	${<}\,0.15$			&	-	&	-	\\
NGC\,55 	&	 $1.981 \pm 0.020$   	&	 3.79 		&	-39.22		&	$\checkmark$	&			&			&	 $0.186 \pm 0.010$  	&		-		&	${<}\,0.37$			&	-	&	-	\\
IC\,342 	&	 $2.24 \pm 0.10$   	&	 56.70 		&	 68.10 		&	$\checkmark$	&			&	$\checkmark$	&	 $0.40 \pm 0.03$  	&		-		&	${<}\,0.89$			&	-	&	-	\\
Circinus 	&	 $2.41 \pm 0.26$ 	&	 213.29 	&	 -65.33		&	$\checkmark$	&			&			&	 $0.68 \pm 0.13$       	&	 J1413.1-6519 	&	 $8.3 \pm 1.2$ 			&	++	&	+	\\
NGC\,2403 	&	 $3.13 \pm 0.06$ 	&	 114.37 	&	 65.59		&	$\checkmark$	&			&			&	 $0.355 \pm 0.012$  	&	 J0737.4+6535 	&	 $1.79 \pm 0.39$	 	&	o	&	o	\\
NGC\,1569 	&	 $3.25 \pm 0.24$ 	&	 67.70 		&	 64.85 		&	$\checkmark$	&			&			&	 $0.70 \pm 0.08$  	&		-		&	${<}\,2.34$			&	-	&	-	\\
M81     	&	 $3.61 \pm 0.20$   	&	148.89		&	 69.07 		&	$\checkmark$	&			&			&	 $0.42 \pm 0.04$  	&		-		&	${<}\,0.32$			&	-	&	-	\\
M82 		&	 $3.53 \pm 0.03$  	&	 148.95 	&	 69.67		&	$\checkmark$	&	$\checkmark$	&	$\checkmark$	&	 $10.41 \pm 0.19\phantom{0}$   	&	 J0955.7+6940     &	 $17.0 \pm 1.0$ &	++	&	+	\\
NGC\,253 	&	 $3.61 \pm 0.03$  	&	 11.90 		&	 -25.29		&	$\checkmark$	&	$\checkmark$	&	$\checkmark$	&	 $5.19 \pm 0.09$ 	&	 J0047.5-2517 	&	 $14.0 \pm 1.0$ 		&	++	&	+	\\
NGC\,4945 	&	 $4.06 \pm 0.77$ 	&	 196.36		&	 -49.47		&	$\checkmark$	&	$\checkmark$	&	$\checkmark$	&	 $1.46 \pm 0.44$	&	 J1305.4-4928 	&	 $16.5 \pm 0.9$ 		&	++	&	+	\\
M83     	&	 $4.79 \pm 0.09$   	&	204.25		&	-29.87		&	$\checkmark$	&	$\checkmark$	&	$\checkmark$	&	 $3.29 \pm 0.10$  	&		-		&	${<}\,2.27$			&	+	&	-	\\
NGC\,3621 	&	 $6.70 \pm 0.37$ 	&	169.57		&	-32.81		&	$\checkmark$	&			&			&	 $0.86 \pm 0.07$  	&		-		&	${<}\,1.72$			&	-	&	-	\\
M101    	&	 $6.76 \pm 0.11$   	&	210.80		&	 54.35 		&	$\checkmark$	&			&			&	 $3.13 \pm 0.22$  	&		-		&	${<}\,0.74$			&	-	&	-	\\
NGC\,6946 	&	 $6.95 \pm 0.38$ 	&	308.71		&	 60.15 		&	$\checkmark$	&			&	$\checkmark$	&	 $3.70 \pm 0.33$  	&		-		&	${<}\,0.36$			&	-	&	-	\\
M106    	&	 $7.54 \pm 0.09$   	&	184.74		&	 47.30 		&	$\checkmark$	&			&			&	 $1.03 \pm 0.08$  	&		-		&	${<}\,0.78$			&	-	&	-	\\
M51     	&	 $8.34 \pm 0.30$   	&	202.47		&	 47.23 		&	$\checkmark$	&			&			&	 $2.83 \pm 0.15$  	&		-		&	${<}\,1.13$			&	-	&	-	\\
NGC1068 	&	 $13.4 \pm 1.8\phantom{0}$ 	&	 40.67	 	&	 -0.01		&			&	$\checkmark$	&	$\checkmark$	&	 $40 \pm 10$	 	&	 J0242.6-0000 	&	 $9.12 \pm 0.68$	 	&	++	&	o	\\
NGC\,2146 	&	 $17.2 \pm 3.2\phantom{0}$ 	&	 94.53	 	&	 78.33		&			&			&			&	 $14.0\pm 5.1\phantom{0}$ 		&	 J0618.1+7819 	&	 $2.49 \pm 0.38$	 	&	o	&	-	\\
NGC\,3424 	&	 $18.8 \pm 3.7\phantom{0}$ 	&	 162.91 	&	 32.89		&			&			&			&	 $0.86 \pm 0.32$    	&	 J1051.6+3253 	&	 $1.32 \pm 0.35$	 	&	o	&	-	\\
NGC\,7059 	&	 $24.7 \pm 4.6\phantom{0}$ 	&	 321.84 	&	 -60.01		&			&			&			&	 $0.67 \pm 0.09$       	&	 J2127.6-5959 	&	 $1.30 \pm 0.34$	 	&	?	&	?	\\
Arp\,299 	&	 $46.8 \pm 3.3\phantom{0}$ 	&	 172.07 	&	 58.52		&			&			&			&	 $97 \pm 14$	 	&	 J1128.2+5831 	&	 $2.00 \pm 0.37$	 	&	o	&	-	\\
Arp\,220 	&	 $80.9 \pm 5.7\phantom{0}$ 	&	 233.70 	&	 23.53		&			&			&			&	 $214 \pm 32\phantom{0}$	 	&	 J1534.7+2331 	&	 $2.99 \pm 0.53$	 	&	o	&	-	\\
\hline

\hline \hline 
\end{tabular}
\tablefoot{The first two columns provide the common name of the galaxy, $d$ is the luminosity distance, R.A. and Dec are the right ascension and declination in ICRS reference system of the center of the associated 4FGL source. The next three columns indicate whether the source meets the selection criterion described in section~\ref{sec: nonGeVDet} (flux relative to that of NGC 253 ${>}\, 1/3.5$) in the Revised MANGROVE Sample \citep{Biteau2021}, the IRAS Revised Bright Galaxy Sample \citep{Sanders2003}, and the sample proposed by \cite{Ackermann2012}. $\dot{M}_*$ is the SFR. 4FGL Name is the name of the associated $\gamma$-ray source in 4FGL-DR4. $\mathcal{F}_\mathrm{2\,GeV}$ is the energy flux at 2\,GeV. The last two columns qualify the reliability of detection by the new $\gamma$-ray observatories, whether pointed (CTAO) or wide field (LHAASO, SWGO): ``++'', ``+'', ``o'', ``-'', ``?'' indicate, respectively, a high degree of reliability, a good degree of reliability, a possibility of detection, an unlikely detection, and that a detailed study is needed before concluding.}
\end{table}
\end{landscape}

\end{document}